\documentclass[aps,pre,reprint,groupedaddress]{revtex4-2}

\usepackage{graphicx}
\usepackage{bm}
\usepackage{amssymb}
\usepackage{amsbsy}
\usepackage{amsmath}
\usepackage{mathtools}
\usepackage{xcolor}
\usepackage{color}
\usepackage{tikz}
\usepackage{hyperref}

\usepackage{epstopdf}

\begin{document}

\def\tr{\textcolor{red}}
\def\tb{\textcolor{blue}} 


\title{Effect of speed fluctuations on the collective dynamics of active disks}


\author{R. Kailasham}
\affiliation{Department of Chemical Engineering, Carnegie Mellon University, Pittsburgh, PA 15213, USA}
\author{Aditya S. Khair}
\email{akhair@andrew.cmu.edu}
\affiliation{Department of Chemical Engineering, Carnegie Mellon University, Pittsburgh, PA 15213, USA}

\date{\today}

\begin{abstract}
Numerical simulations are performed on the collective dynamics of active disks, whose self-propulsion speed ($U$) varies in time, and whose orientation evolves according to rotational Brownian motion. Two protocols for the evolution of speed are considered: (i) a deterministic one involving a periodic change in $U$ at a frequency $\omega$; and (ii) a stochastic one in which the speeds are drawn from a power-law distribution at time-intervals governed by a Poissonian process of rate $\beta$. In the first case, an increase in $\omega$ causes the disks to go from a clustered state to a homogeneous one through an apparent {phase-transition}, provided that the direction of self-propulsion is allowed to reverse. Similarly, in the second case, for a fixed value of $\beta$, the extent of cluster-breakup is larger when reversals in the self-propulsion direction are permitted. Motility-induced phase separation of the disks may therefore be avoided in active matter suspensions in which the constituents are allowed to reverse their self-propulsion direction, immaterial of the precise temporal nature of the reversal (deterministic or stochastic). Equally, our results demonstrate that phase separation could occur even in the absence of a time-averaged motility of an individual active agent, provided that the rate of direction reversals is smaller than the orientational diffusion rate.
\end{abstract}

\maketitle






\section{\label{sec:intro}Introduction}
Self-propelled entities harness energy from their surroundings or use an inner motive power to navigate the environment around them. These active objects could either be living matter, like flagellate microorganisms~\cite{Wan2018} or epithelial cells moving during the wound-healing process~\cite{Camley2017}; or non-living, like droplets moving in response to a chemical gradient or Marangoni flows~\cite{Izri2014,Meredith2020,Hokmabad2021,Suda2021}, or programmable robots.~\cite{Nava2018,Horvath2023} Abstracting away the physical mechanism responsible for their sustained self-propulsion, a commonly used reduced-order model to interpret the dynamics of such particles is the Active Brownian Particles~\cite{Ebbens2010,TenHagen2011,Bechinger2016,Marchetti2016,Zeitz2017,Callegari2019,Zottl2023} (ABP) framework, in which Langevin equations are specified for the position and orientation of the particle. The particle is assumed to possess a \textit{constant} speed of self-propulsion, while its orientation evolves according to a Gaussian white noise process.~\cite{TenHagen2009,Ebbens2010,TenHagen2011} While the conventional ABP model assumes a constant speed of self-propulsion, fluctuations in the speed of self-propelled particles are relevant in modeling the dynamics of biological~\cite{Bazazi2011,Zaburdaev2011,Otte2021} and non-biological active agents.~\cite{Grosmann2012} For instance, a recent investigation on the chaotic dynamics of an autophoretic disk in a quiescent fluid establishes that the instantaneous speed and orientation of a disk fluctuate over the course of its motion.~\cite{Kailasham2022} The dynamics of a single active disk moving with a fluctuating self-propelled speed has been explored in detail~\cite{Lauga2011,Peruani2007}, but that of a collection of interacting active disks with speed fluctuations remain less so.~\cite{Grosmann2012,Romanczuk2012,Mahault2023} We seek to address this question in the present work.

Computational models incorporating the minimal ingredients of a constant self-propulsion speed and repulsive interactions are known to predict the clustering, or motility-induced phase separation~\cite{Fily2012,Stenhammar2014,Cates2015,Digregorio2018} (MIPS), of active entities observed in experiments~\cite{Buttinoni2013,Bialke2015}. A natural question, and one which forms the crux of this paper, is how do fluctuations in the self-propulsion speed affect this phase-separation process? Two cases of speed variations are considered herein: (i) deterministic, in which the speed varies periodically in time with an angular frequency $\omega$, as in the reciprocal swimmer (henceforth RS) model studied by~\citet{Lauga2011}; and (ii) stochastic, in which the speeds are sampled from an arbitrary distribution, with the time of update governed by a Poissonian process with the rate constant $\beta$~\cite{Peruani2007,Callegari2019}, as proposed by~\citet{Peruani2007} (henceforth PM). For each of the two update procedures, we also consider sub-cases which permit or disallow negative values for the self-propulsion speed $U$, i.e., with and without reversals in the direction of self-propulsion. In the deterministic speed-update case, a central finding is that a collection of sterically repulsive RS disks undergoing periodic reversals in the swimming direction exhibit phase-separation at small values of $\omega$, despite the absence of a mean (in the time-averaged sense) swimming speed. Thus, this is an example of MIPS without (net) motility. At large values of $\omega$, however, there is no phase separation. Notably, this order-disorder transition vanishes when directional reversals are disallowed. A similar trend is also observed for the stochastic speed-update rule: directional reversals amplify the impact of speed fluctuations in driving the system to a homogeneous phase. Additionally, in both cases the dynamics of the active suspension are characterized using their mean square displacement (MSD) and velocity autocorrelation (VAC).

The rest of the paper is organized as follows: the governing equations for a single disk moving with deterministic and stochastic speed fluctuations are presented in Section~\ref{sec:num_alg}, along with the details of the numerical algorithm used to simulate a collection of sterically repulsive disks. Results for the deterministic and stochastic speed-fluctuation protocol are presented in Secs.~\ref{sec:det_speed} and~\ref{sec:prob_speed}, respectively, and we conclude in Sec.~\ref{sec:concl}. The dynamics of PM disks whose speeds are drawn from a uniform random distribution is discussed in Appendix~\ref{sec:unif_pm}. Simulation videos for the various speed-update protocols are available in the Supplementary Material, {as well as a comparison of the results obtained using two simulation algorithms}.

\section{\label{sec:num_alg}Numerical algorithm}

\begin{figure}[t]
\centering
\includegraphics[width=3.1in,height=!]{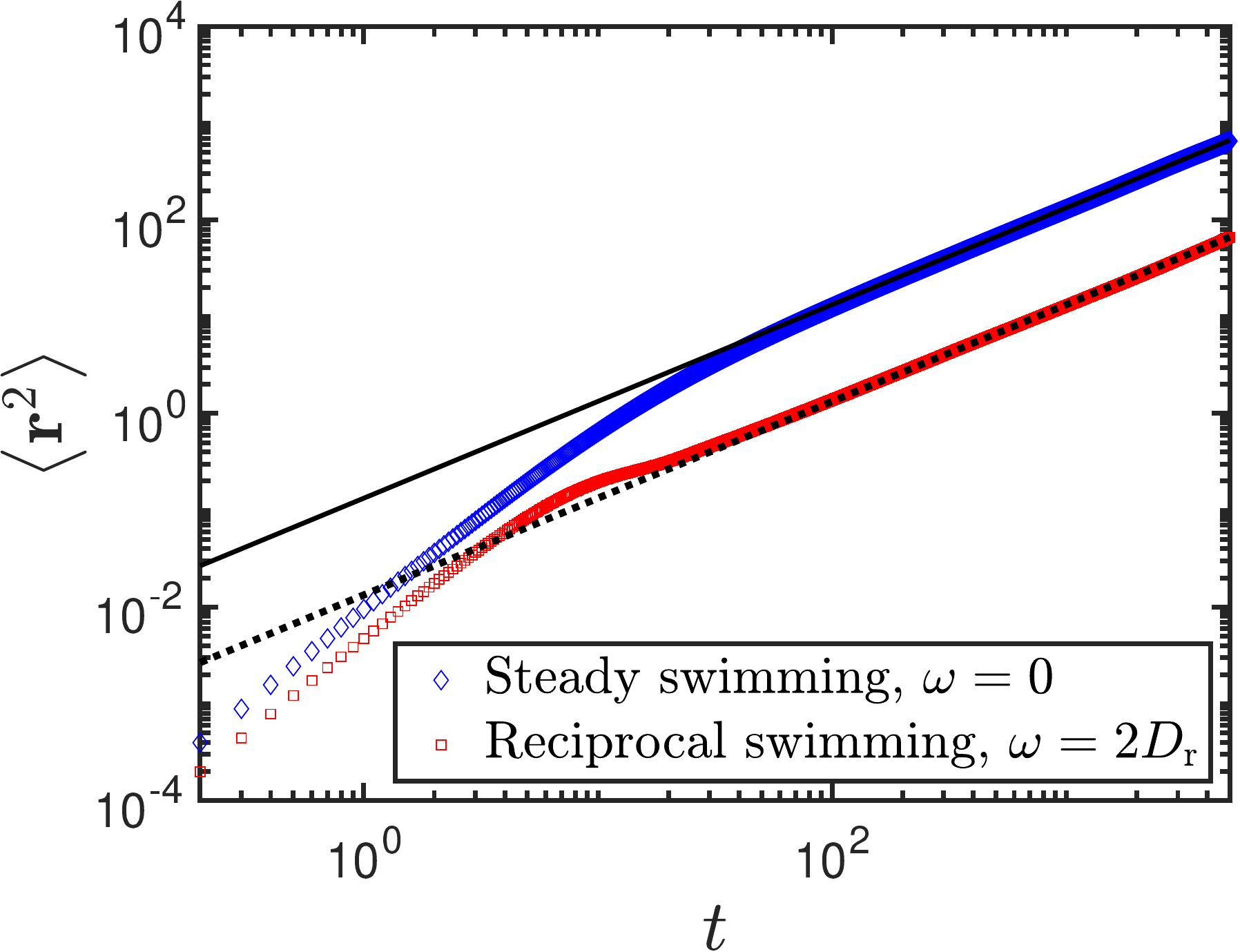}
\caption{Mean square displacement of a single RS disk. The broken and solid lines correspond to equations~(\ref{eq:d0_oscill}) and~(\ref{eq:d0_stead}), respectively, while the symbols represent results from numerical simulations. The MSD was averaged over 300 independent realizations, for disks with $U_{0}=0.1$ and $D_{\text{r}}=0.15$.}
\label{fig:msd_oscill}
\end{figure}

\begin{figure*}[t]
\centering
\includegraphics[width=6in,height=!]{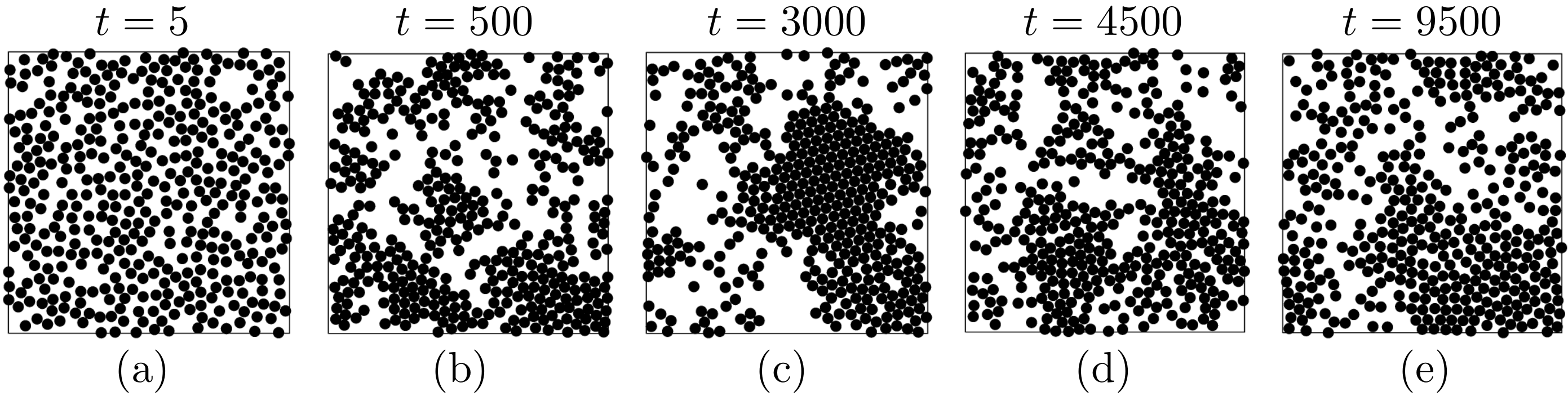}
\caption{Simulation snapshots for a collection of RS disks whose speeds oscillate at an angular frequency of $\omega=0.5\,D_{\text{r}}$, recorded at (a) $t=5$, (b) $t=500$, (c) $t=3000$, (d) $t=4500$, and (e) $t=9500$. The rotational diffusion constant is $D_{\text{r}}=10^{-3}$, and the size of the system is $N=400,\,L=10$.}
\label{fig:schem_five_times}
\end{figure*}

We first consider a single disk whose state at any discrete time instance $i$ is characterized by its velocity $\bm{U}(i)$, and orientation $\theta(i)$ measured with respect to the positive $x$-axis. The $x$- and $y$-components of the instantaneous position are updated using the explicit Euler method as follows
\begin{equation}\label{eq:pos_update}
\begin{split}
r_{x}(i+1)=r_{x}(i)+U_{x}(i)\Delta t;\\
r_{y}(i+1)=r_{y}(i)+U_{y}(i)\Delta t,
\end{split}
\end{equation}
where $\Delta t$ is the timestep width, $U_{x}(i)=U(i)\cos\theta(i);\,U_{y}(i)=U(i)\sin\theta(i)$, with $U(i)$ denoting the instantaneous speed of the disk, and the absolute time related to the discrete time as $t_i=i\Delta t$. The time-evolution of the orientation follows a diffusive process, 
\begin{equation}
\theta(i+1)=\theta(i)+\sqrt{2D_{\text{r}}}\eta(t),
\end{equation}
where $D_{\text{r}}$ is the rotational diffusion constant, and $\eta(t)$ is Gaussian white noise with zero mean and a variance of $\Delta t$. 

Two cases of speed evolution are considered: deterministic and stochastic. In the former case, the magnitude of the speed varies periodically in time~\cite{Lauga2011}, as $U(i)=U_0\cos\left(\omega i\Delta t\right)$. Hence, there is no net motility over one time period, $2\pi/\omega$, of the motion. In the latter case of stochastic speed fluctuations, we adopt the recipe prescribed by PM for updating the instantaneous speed, in which $U(i)$ is sampled from a power law distribution $P(U)\sim U^{-\alpha}$ in the window $[U_0,nU_0]$. The mean $\left<U\right>$ and variance $\sigma^2=\left(\left<U^2\right>-\left<U\right>^2\right)$ of the stationary distribution are defined as follows
\begin{equation}\label{eq:av_U_pwlaw}
\left<U\right>=\left(\dfrac{\alpha-1}{\alpha-2}\right)U_{0}\left[\dfrac{1-\left(1/n\right)^{\alpha-2}}{1-\left(1/n\right)^{\alpha-1}}\right],
\end{equation}
\begin{equation}\label{eq:var_U_pwlaw}
\sigma^2=\left(\dfrac{\alpha-1}{\alpha-3}\right)U^2_{0}\left[\dfrac{1-\left(1/n\right)^{\alpha-3}}{1-\left(1/n\right)^{\alpha-1}}\right]-\left<U\right>^2.
\end{equation}
The frequency of updating the instantaneous speed is governed by a Poissonian process of rate $\beta$. The speed of all particles are set to an initial value of $U_0$. The time-step $i^{(\text{up})}$ for the next update for the speed is determined stochastically by~\cite{Callegari2019} 
\begin{equation}
i^{\text{(up)}}=-\dfrac{\ln(s)}{\beta\Delta t}, 
\end{equation}
where $s$ is drawn randomly from a uniform distribution in $[0,1]$. The instantaneous speed retains its value between updates. The special case of $\beta=0$ implies an absence of speed updates. A key dimensionless parameter, $\gamma$, is the ratio of the timescales for speed and orientation fluctuations. It is defined as $\gamma\equiv\omega/D_{\text{r}}$ for RS disks and $\gamma\equiv\beta/D_{\text{r}}$ for PM disks. The results for the deterministic and stochastic self-propulsion schemes are presented in subsequent sections. 

The procedure used to simulate a collection of $N$ disks with steric repulsion, confined in a square box of size $L$ with periodic boundary conditions, is described next. The area fraction of disks in the box is $\phi=N\pi\,d^2/4L^2$, with $d$ denoting the diameter of the disk. The finite size of the disks is accounted for by incorporating a steric repulsion that operates only when two disks come in contact. We resolve inter-particle overlaps through the Heyes-Melrose algorithm~\cite{Heyes1993}, as implemented in ref.~\citenum{Foss2000}. The particles are allowed to evolve in time according to eq.~(\ref{eq:pos_evol}). At each timestep, the algorithm checks for overlaps, defined to occur when the centre-centre separation between two particles $j$ and $k$ is less than their diameter $d$. The positions of each particle in a contact pair are reset harmonically by an amount proportional to the extent of overlap, so that the disks just touch each other. The total velocity on the $j^{\text{th}}$ particle due to overlaps is evaluated as
\begin{equation}\label{eq:hm_vel}
\bm{v}^{\text{HM}}_{j}=\dfrac{1}{\Delta t}\sum_{k}^{n_{o}}K\left(r_{jk}-d\right)\Theta\left(d-r_{jk}\right)\hat{\bm{r}}_{jk},
\end{equation}
where $\hat{\bm{r}}_{jk}=\bm{r}_{jk}/r_{jk}$ is the unit vector along the line joining the particle centres, $\Theta$ denotes the Heaviside function, the stiffness $K=0.5$, and $n_{o}$ denotes the number of overlapping particles in the neighborhood of the $j^{\text{th}}$ particle. The position of the $j^{\text{th}}$ particle is then reset by an amount $\bm{v}^{\text{HM}}_{j}\Delta t$. The procedure is repeated until all particle overlaps are resolved at a given instant of time. The positional update rule therefore reads as follows
\begin{equation}\label{eq:pos_evol}
\bm{r}_{j}(t_i+\Delta t)=\bm{r}_{j}(t_i)+\bm{U}_{j}(t_i)\Delta t+\bm{v}^{\text{HM}}_{j}(t_i)\Delta t; j=1,2,\cdots,N.
\end{equation}
The instantaneous velocity of the $j^{\text{th}}$ disk is defined as $\bm{v}_{j}(t_i)=\left[\bm{r}_{j}(t_i+\Delta t)-\bm{r}_{j}(t_i)\right]/\Delta t$, with the understanding that $\bm{v}\equiv\bm{U}$ for an isolated disk in a box, where $\bm{U}$ is defined in eq.~(\ref{eq:pos_update}). In numerical implementation of the Heyes-Melrose algorithm, we employ a finite softness~\cite{Heyes1993} of $0.01d$, while determining particle overlaps. A value of $d=0.4$ has been used for all the disks considered in the present work, {unless specified otherwise}. 
The minimum image convention~\cite{FrenkelSmit2001} is employed to compute the interactions between particles, which assumes that each disk interacts only with its nearest image, and not an infinite number of copies of itself. 

A value of $U_{0}=0.1$ has been used in all simulations, unless specified otherwise. The values of $\Delta t$ are chosen such that the displacement over a single time-step does not exceed the diameter of a single disk. A value of $\Delta t=10^{-1}$ is found to suffice for the case of RS disks with and without directional reversals, since their self-propulsion speeds do not exceed $0.1$ in magnitude. For the case of PM disks without direction reversal, the speeds are drawn from a power-law distribution such that $U\in\left[0.1,1000\right]$. Given the broad-range of allowable values for $U$, a timestep width of $\Delta t=10^{-4}$ has been used for simulations of PM disks. For PM disks undergoing a reversal in the swimming direction, the speeds are sampled from a uniform random distribution in the interval $[-0.1,0.1]$ and $\Delta t=10^{-1}$ is used. Additional simulations have been performed for select-cases and we observe that the results presented below are not changed qualitatively as $\Delta t$ is varied.

The P\'{e}clet number associated with a single disk may be defined~\cite{Redner2013,Stenhammar2014} on the basis of $U_{0}$ as $Pe=3U_{0}/dD_{\text{r}}$. For all investigations of MIPS, an area fraction of $\phi=0.5$ and $D_{\text{r}}=10^{-3}$ are chosen, such that the state point $\left(\phi=0.5,Pe=750\right)$ lies {deep} in the phase-separated region for conventional ABPs undergoing MIPS~\cite{Stenhammar2014,Speck2015}. {For an area fraction of $\phi=0.5$, a value of $Pe=100$ is sufficient to observe phase separation in a system of conventional ABPs interacting via steric repulsion~\cite{Stenhammar2014}. The RS disks (with and without direction reversal) and PM disks with direction reversal considered in our study undergo periodic fluctuations in speed, and a value of $Pe=750$ is defined on the peak value ($U_{0}=0.1$) of the speed. Therefore, the instantaneous P\'{e}clet number varies from a largest value of 750 to small values, even reaching zero, over a time period of speed oscillation. We have therefore chosen to operate at a value of $Pe$ that is far from the phase-boundary in the $Pe-\phi$ parameter space recorded for conventional ABPs.}


{For systems with $N\leq1600$, eq.~(\ref{eq:pos_evol}) is integrated using MATLAB~\cite{Kailasham_PM_and_RS_2023}. To test the robustness of our predictions at various system sizes, we performed large scale simulations with $N=\mathcal{O}(10^4)$ number of particles using a custom-written code in HOOMD-blue~\cite{Anderson2020}. Steric repulsion is implemented in HOOMD-blue through a hard-sphere potential that is non-zero only for inter-particle separations below the particle diameter. The results from both the implementations provide comparable results (see Fig.~S1 in Supplementary Material), and we have used HOOMD-blue for larger scale simulations due to its superior computational performance over the MATLAB code.}

\section{\label{sec:det_speed}Deterministic speed evolution}

\begin{figure*}[t]
\begin{center}
\begin{tabular}{c c c}
\includegraphics[width=2.3in,height=!]{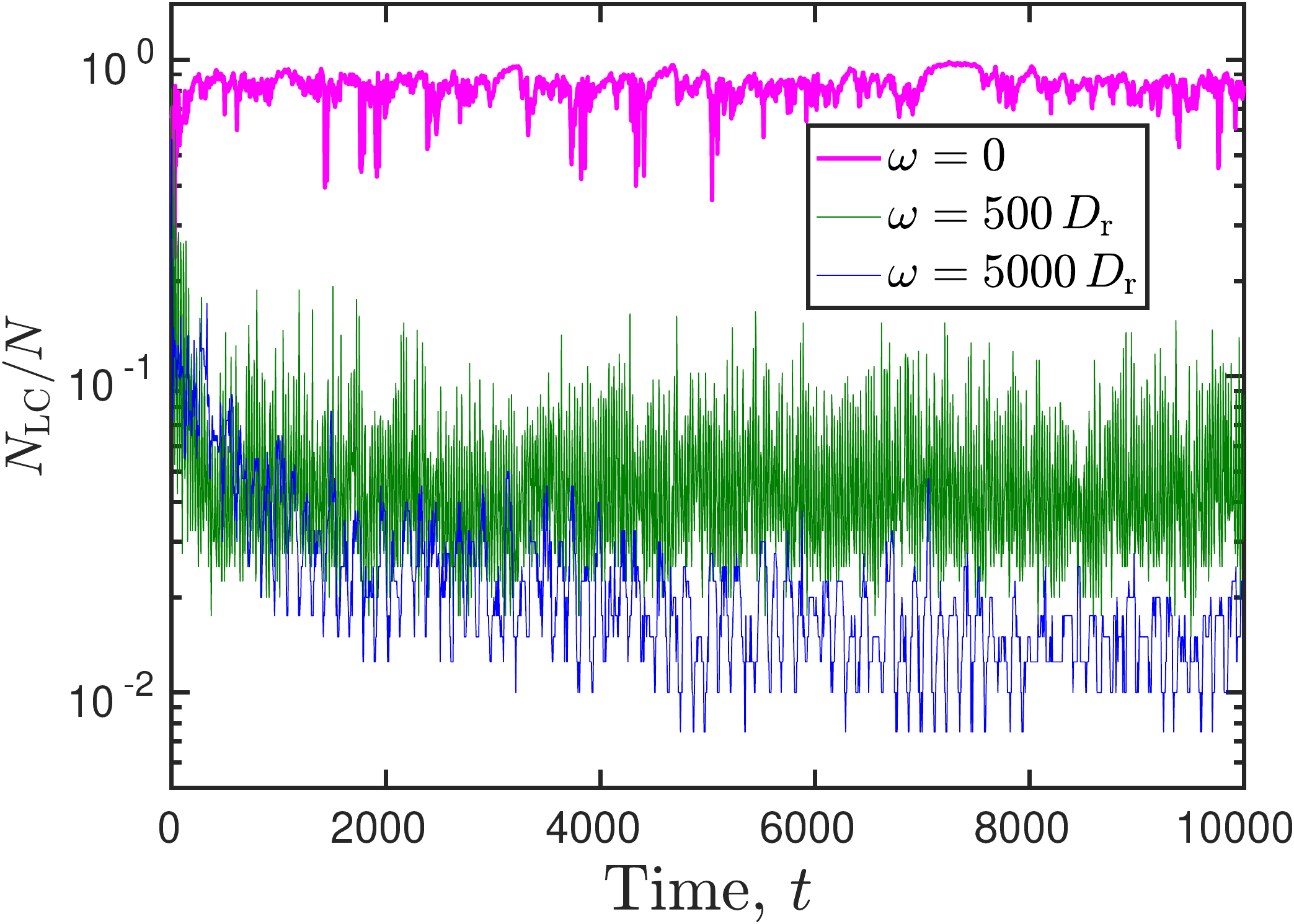}&
\includegraphics[width=2.3in,height=!]{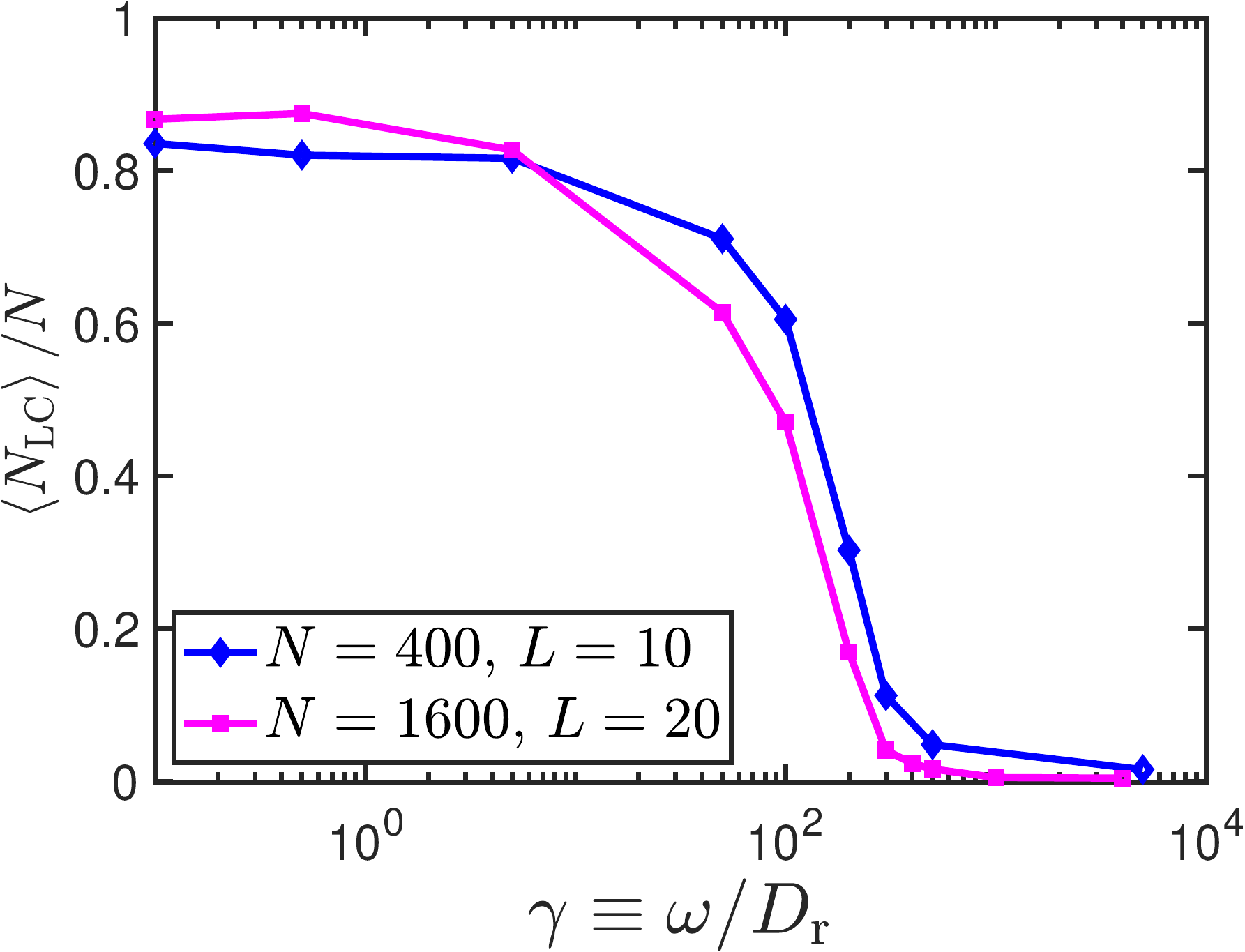}&
\includegraphics[width=2.3in,height=!]{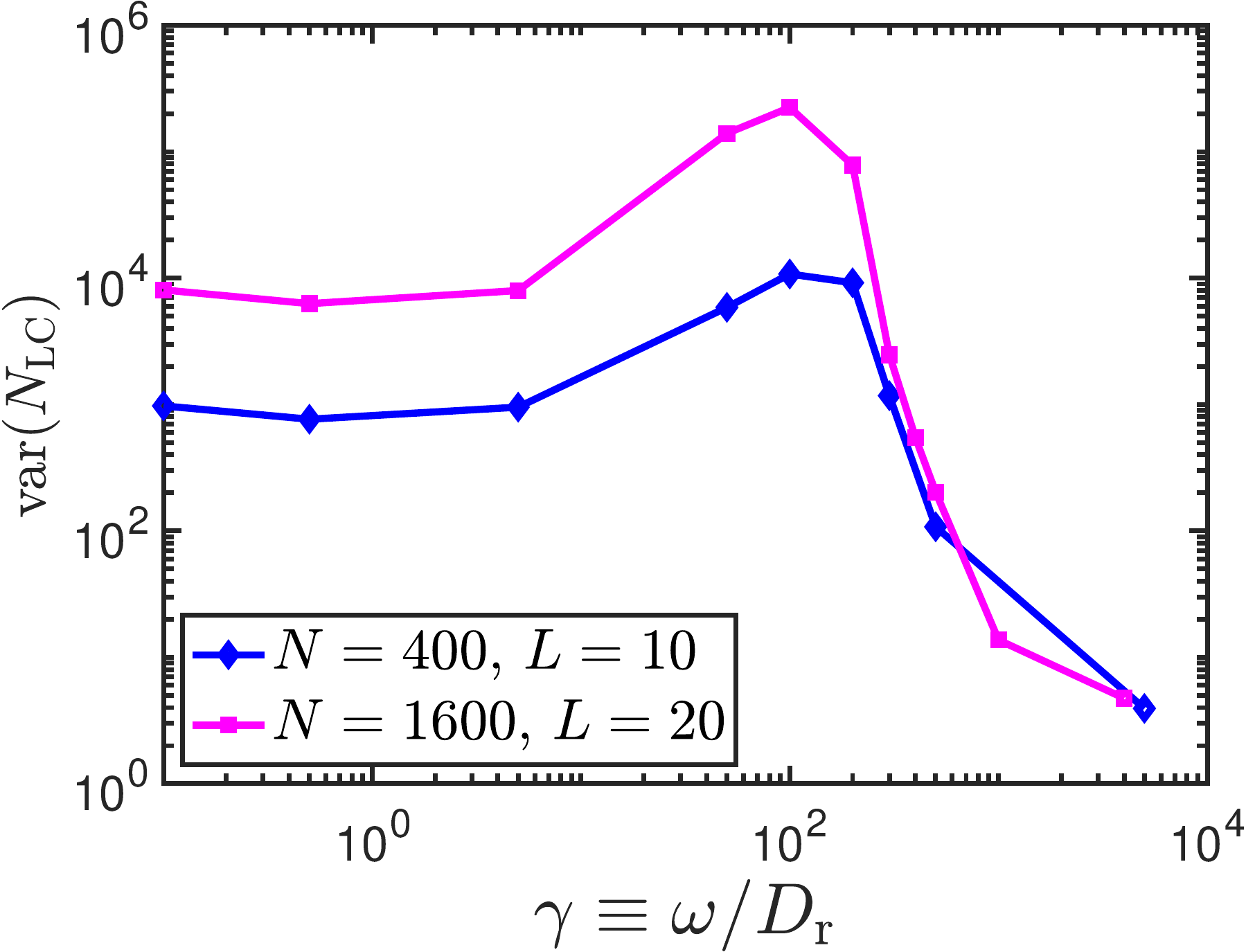}\\
(a)&(b)&(c)\\
\end{tabular}
\end{center}
\caption{(a) Instantaneous size of the largest cluster ($N_{\text{LC}}$) normalized by the number of RS disks in the box, (b) average size of the largest cluster and (c) the standard deviation in $N_{\text{LC}}$ for various values of the oscillation frequency. The rotational diffusion constant is $D_{\text{r}}=10^{-3}$, and two different system sizes are considered at a fixed area fraction of $\phi=0.5$. The cluster statistics are evaluated once $N_{\text{LC}}$ has attained a steady state, which in (a) is observed to occur after $t\geq6000$ for a box with $N=400,\,L=10$.}
\label{fig:oscill_mips}
\end{figure*}

\begin{figure*}[t]
\centering
\includegraphics[width=6in,height=!]{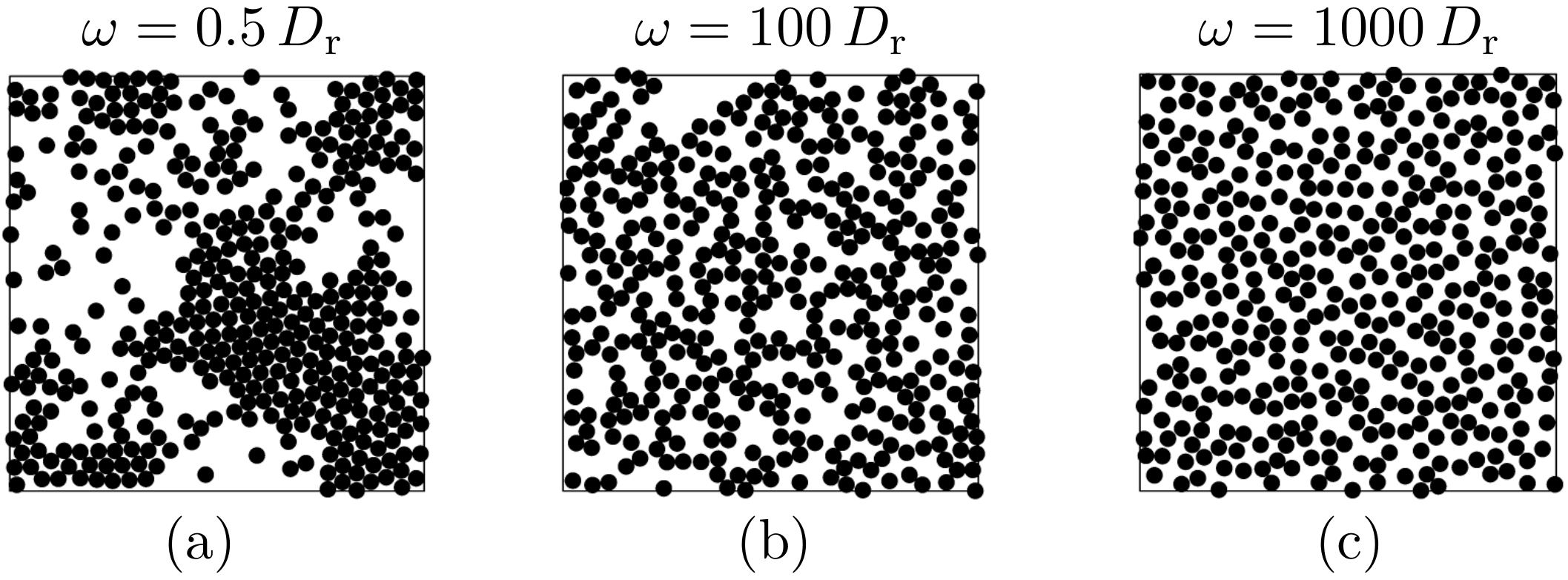}
\caption{Simulation snapshots at the same time-instance ($t=8750$) for a collection of RS disks whose speeds oscillate at an angular frequency of (a) $\omega=0.5\,D_{\text{r}}$, (b) $\omega=100\,D_{\text{r}}$ and (c) $\omega=1000\,D_{\text{r}}$. The rotational diffusion constant is $D_{\text{r}}=10^{-3}$, and the size of the system is $N=400,\,L=10$.}
\label{fig:schem_three_freq}
\end{figure*}

Lauga~\cite{Lauga2011} has derived the following analytical expression for the long-time diffusivity of a single RS disk moving in an unbounded fluid 
\begin{equation}\label{eq:d0_oscill}
D^{\text{L}}_{0}\equiv\lim_{t\to\infty}\dfrac{1}{4t}\left<\bm{r}^2(t)\right>=\dfrac{1}{4}\dfrac{U^2_{0}D_{\text{r}}}{D^2_{\text{r}}+\omega^2},
\end{equation}
such that as $t\to\infty$, the two-dimensional mean square displacement varies as $\left<\bm{r}^2\right>\sim4D^{\text{L}}_{0}t$. For the the special case of a steady swimmer whose speed does not oscillate in time ($\omega=0$), the classical ABP result~\cite{Bechinger2016,Marchetti2016} is recovered, namely
\begin{equation}\label{eq:d0_stead}
D^{\text{L}}_{0}(\omega=0)=\dfrac{U^2_{0}}{2D_{\text{r}}}.
\end{equation} 
Interestingly, 
\begin{equation}
\lim_{\omega\to0}D^{\text{L}}_{0}(\omega)=\dfrac{1}{2}D^{\text{L}}_{0}(\omega=0),
\end{equation}
i.e., the long-time diffusivity of a reciprocating swimmer in the zero-frequency limit is not the same as that for a swimmer moving with a constant speed. This is analogous to a result in Taylor dispersion, in which the dispersion coefficient in oscillatory Poiseuille flow in the limit of zero oscillation frequency is half that in steady flow.~\cite{Bowden1965,Watson1983} The essential feature of~(\ref{eq:d0_oscill}) is that even though the disk experiences no net motility due to its reciprocal motion, it exhibits long-time diffusion, albeit with a reduced diffusion constant compared to the steady swimmer.

In fig.~\ref{fig:msd_oscill}, the numerically evaluated MSD for a single RS disk is compared against Lauga's~\cite{Lauga2011} analytical results for their long-time behavior. The excellent agreement between the two sets of results establishes the validity of our numerical procedure for simulating active disks whose self-propulsion speeds oscillate periodically in time.

We next examine the dynamics of a collection of RS disks interacting purely through steric repulsions, with a view to study the effect of the angular frequency $\omega$ on MIPS. MIPS has been commonly likened to liquid-gas phase separation~\cite{Bialke2013,Stenhammar2013,Levis2017,Caporusso2020}, in which a large cluster of disks dynamically exchanges particles with other smaller clusters. {In Fig.~\ref{fig:schem_five_times}, the time-evolution of configurations for a collection of RS disks that interact through steric repulsions and whose speeds oscillate at a frequency of $\omega=0.5\,D_{\text{r}}$ is plotted. It is useful to quantify the collective dynamics illustrated in Fig.~\ref{fig:schem_five_times} through the size of the largest cluster, $N_{\text{LC}}$.} In Fig.~\ref{fig:oscill_mips}~(a), the size of the largest cluster is plotted as a function of time for several values of the oscillating frequency.  In the absence of speed variation, the majority of the disks in the box belong to the largest cluster, and the size of this cluster is seen to fluctuate {in time} about a mean value (Supplementary Video 1). As the oscillating frequency is increased, however, a homogeneous distribution of disks is the preferred configuration (Supplementary Videos 2-4). This trend is evident from Fig.~\ref{fig:oscill_mips}~(b), in which the mean cluster size is seen to vary in a sigmoidal fashion as a function of the speed-oscillation frequency. {A similar suppression of MIPS has also been observed by~\citet{Omar2023} in a mixture of passive and active disks with periodic (and deterministic) speed fluctuation, as the frequency of speed fluctuation is increased.} As shown in Fig.~\ref{fig:oscill_mips}~(c), the fluctuations in the cluster size show an abrupt increase at a ``critical" frequency $\omega\approx100\,D_{\text{r}}$. {In Fig.~\ref{fig:schem_three_freq}, representative snapshots at the same time-instance and corresponding to three different frequencies are plotted, to illustrate the decrease in cluster size with frequency.}

\begin{figure}[t]
\begin{center}
\begin{tabular}{c}
\includegraphics[width=3in,height=!]{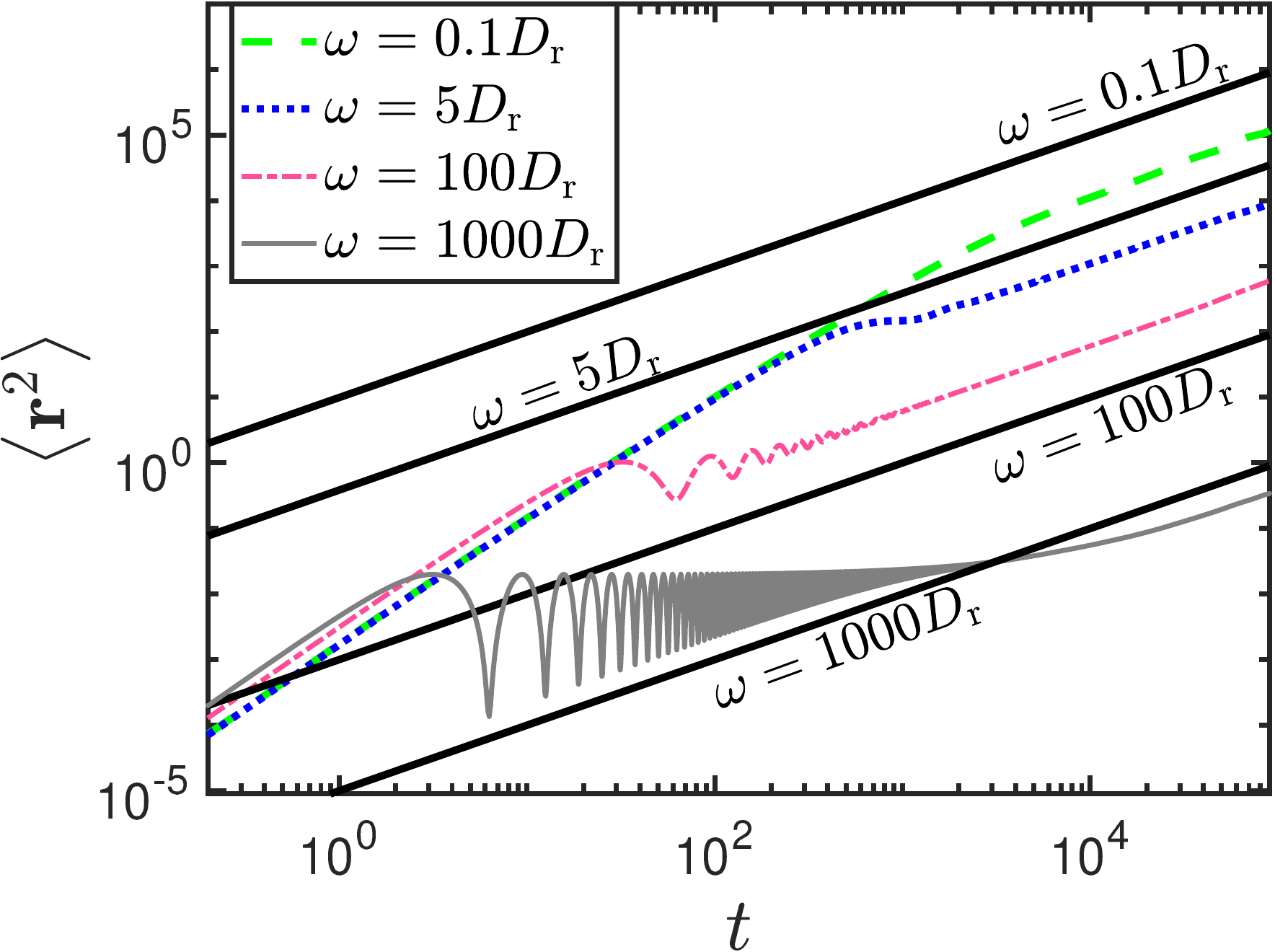}\\
(a)\\
\includegraphics[width=3in,height=!]{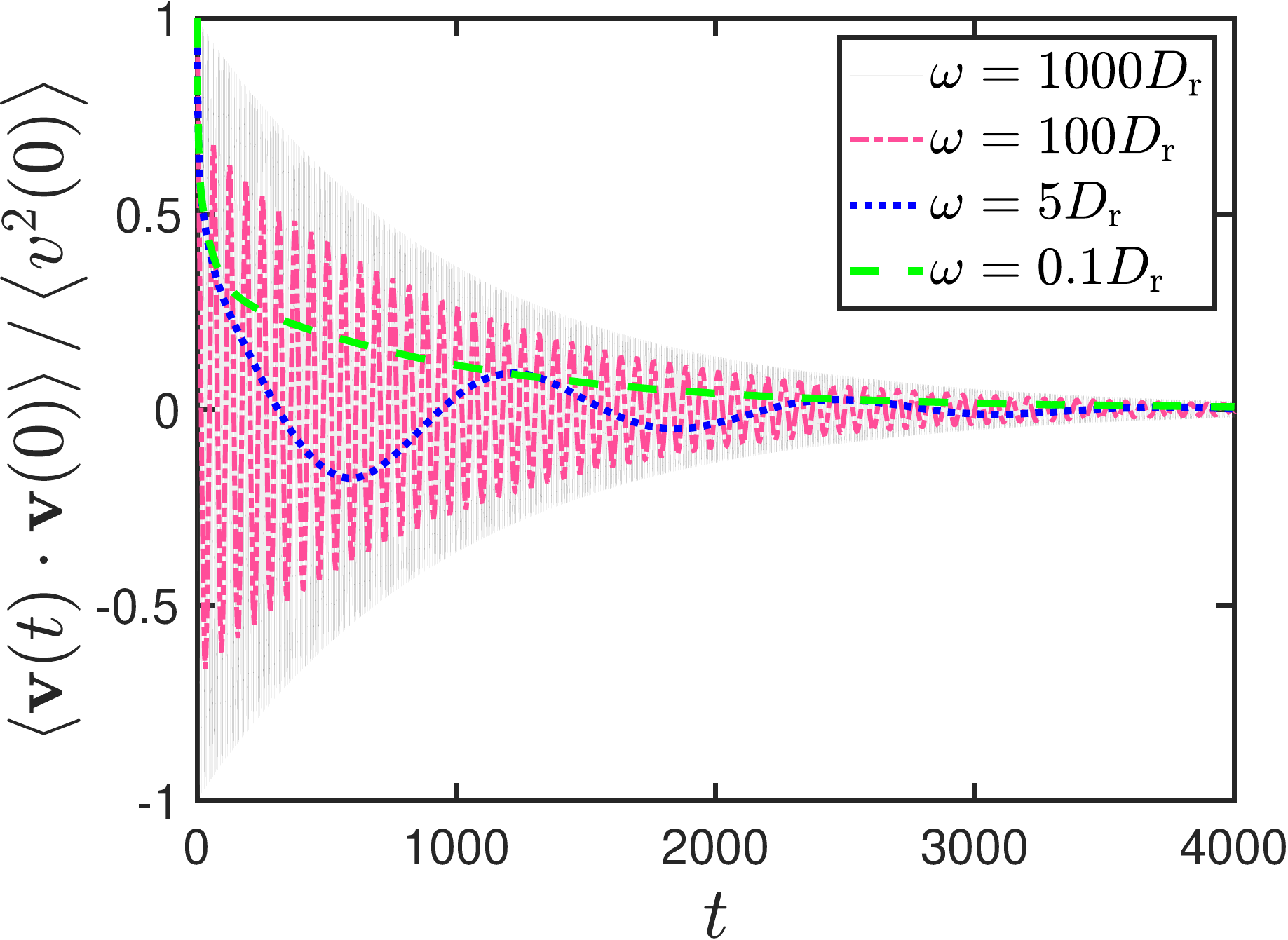}\\
(b)
\end{tabular}
\end{center}
\caption{(a) Mean square displacement and (b) velocity autocorrelation for a collection of RS disks at a fixed area fraction of $\phi=0.5$ and various angular frequencies. The model parameters are $U_0=0.1$ and $D_{\text{r}}=10^{-3}$. {The solid lines in (a) represent the mean square displacement for an isolated RS disk, $\left<\bm{r}^2\right>=4D^{\text{L}}_{0}t$, with $D^{\text{L}}_{0}$ given by eq.~(\ref{eq:d0_oscill}).}}
\label{fig:oscill_msd_vac}
\end{figure}

In Fig.~\ref{fig:oscill_msd_vac}, the MSD and VAC for the collection of RS disks at three different angular frequencies are plotted, for the same area fraction ($\phi=0.5$) considered in Fig.~\ref{fig:oscill_mips}. At the lowest frequency considered in the figure, the long-time diffusion constant is lower than the theoretical prediction for a collection of non-interacting, or phantom, RS disks~(\ref{eq:d0_oscill}). At the critical frequency $\omega\approx100\,D_{\text{r}}$, however, the trend is reversed, with the sterically repulsive disks possessing a larger MSD than their phantom counterparts. For the largest value of $\omega$ considered in the figure, the MSD transiently exceeds the theoretical result, before crossing over to a lower value. We therefore observe that the long-time diffusion constant for a collection of disks on either side of the critical frequency is lower than that for a single disk. The velocity autocorrelation resembles a damped exponential at all the four frequencies.

{It is instructive to compare the dynamics of RS disks against prior simulations of ABPs~\cite{Bialke2012,Redner2013} which differ from the present work in two additional ways besides the non-constancy of the self-propulsion speed. Firstly, ref.~\citenum{Redner2013} accounts for translational diffusion arising from thermal noise, which is ignored in the present work. Secondly, an area fraction of $\phi=0.6$ is considered in their study, and the disks phase separate into an ``active solid" phase and a dilute phase at large enough values of $Pe$. We reiterate that the MSD in Fig.~\ref{fig:oscill_msd_vac}~(a) is an ensemble-average over all the disks in the box, and that these disks move between the dilute and dense phases over the course of the simulation. For the system considered in ref.~\citenum{Redner2013}, the dynamics of a tagged particle in the dilute phase and the active solid cluster differ markedly. The MSD of the free particle (in dilute phase) follows the diffusive-ballistic-diffusive pattern typical of ABPs whose motion is influenced by thermal effects~\cite{Bechinger2016,Marchetti2016,Zeitz2017}. The MSD of the tagged particle in the solid phase however, exhibits an early time $\sim t^{1/2}$ behavior, followed by a $\sim t^{3/2}$ regime before transitioning to long-time diffusion. The MSD at the lowest frequency examined in Fig.~\ref{fig:oscill_msd_vac}~(a) exhibits early time ballistic motion followed by a transition to long-time diffusion, and is significantly different from that reported in ref.~\citenum{Redner2013}. The absence of early-time diffusion can be attributed to the lack of thermal noise in the present work, and the absence of subdiffusive motion could be due to the lower area fraction used in our work. Examining the interplay between area fraction and speed fluctuations on the collective dynamics of active disks is an interesting problem for future work, but is beyond the scope of this paper in which we consider a single area fraction.}

\begin{figure}[t]
\centering
\includegraphics[width=3.1in,height=!]{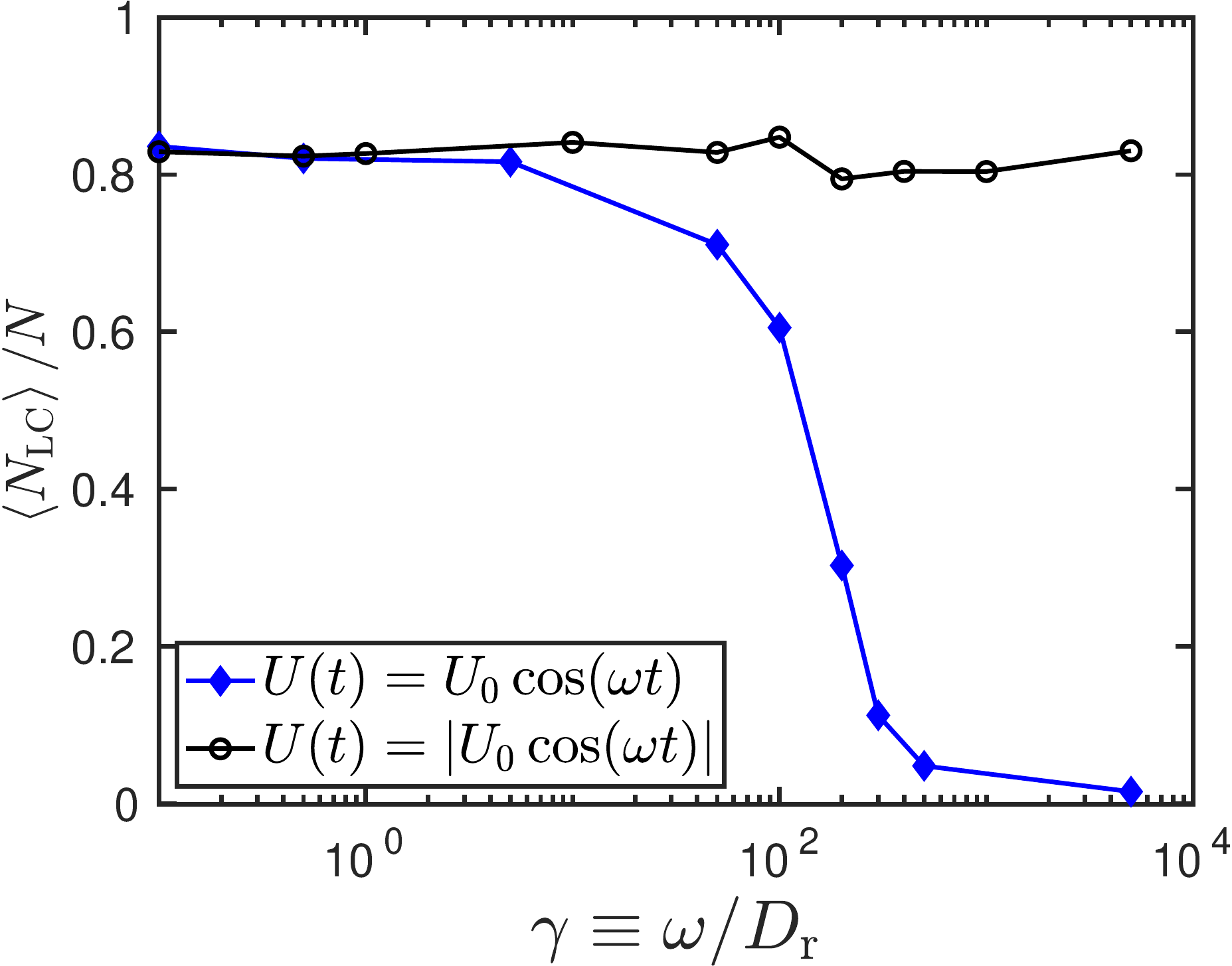}
\caption{Average cluster size as a function of the scaled angular frequency, for a collection of $N=400$ disks in a periodic box of $L=10$. The filled and hollow symbols correspond to RS disks with and without speed reversals, respectively. The model parameters are $U_0=0.1$ and $D_{\text{r}}=10^{-3}$.}
\label{fig:rs_eff_dir_rev}
\end{figure}

We next investigate an alternative deterministic protocol in which the swimming speed of the RS disks varies as $U(t)=|U_{0}\cos(\omega t)|$. In this case, the self-propulsion speed of the disks varies periodically in time but does not change direction, i.e., does not take negative values. In Fig.~\ref{fig:rs_eff_dir_rev}, the effect of this protocol on the cluster statistics is plotted as a function of the scaled frequency. Clearly, it is observed that the order-disorder transition does not occur in the absence of directional reversal for a deterministic speed-evolution protocol.

\begin{figure}[t]
\begin{center}
\begin{tabular}{c}
\includegraphics[width=3in,height=!]{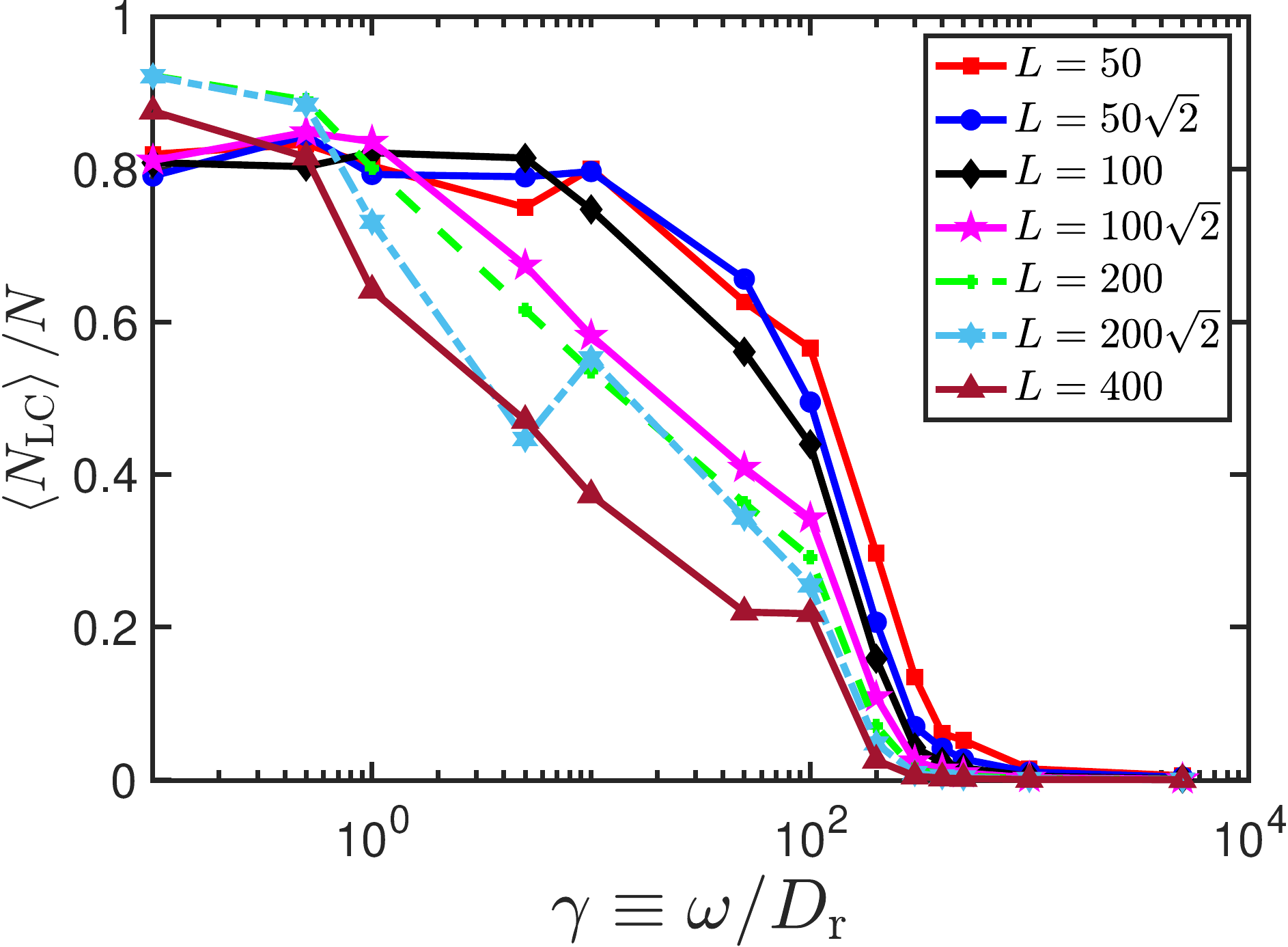}\\
(a)\\
\includegraphics[width=3in,height=!]{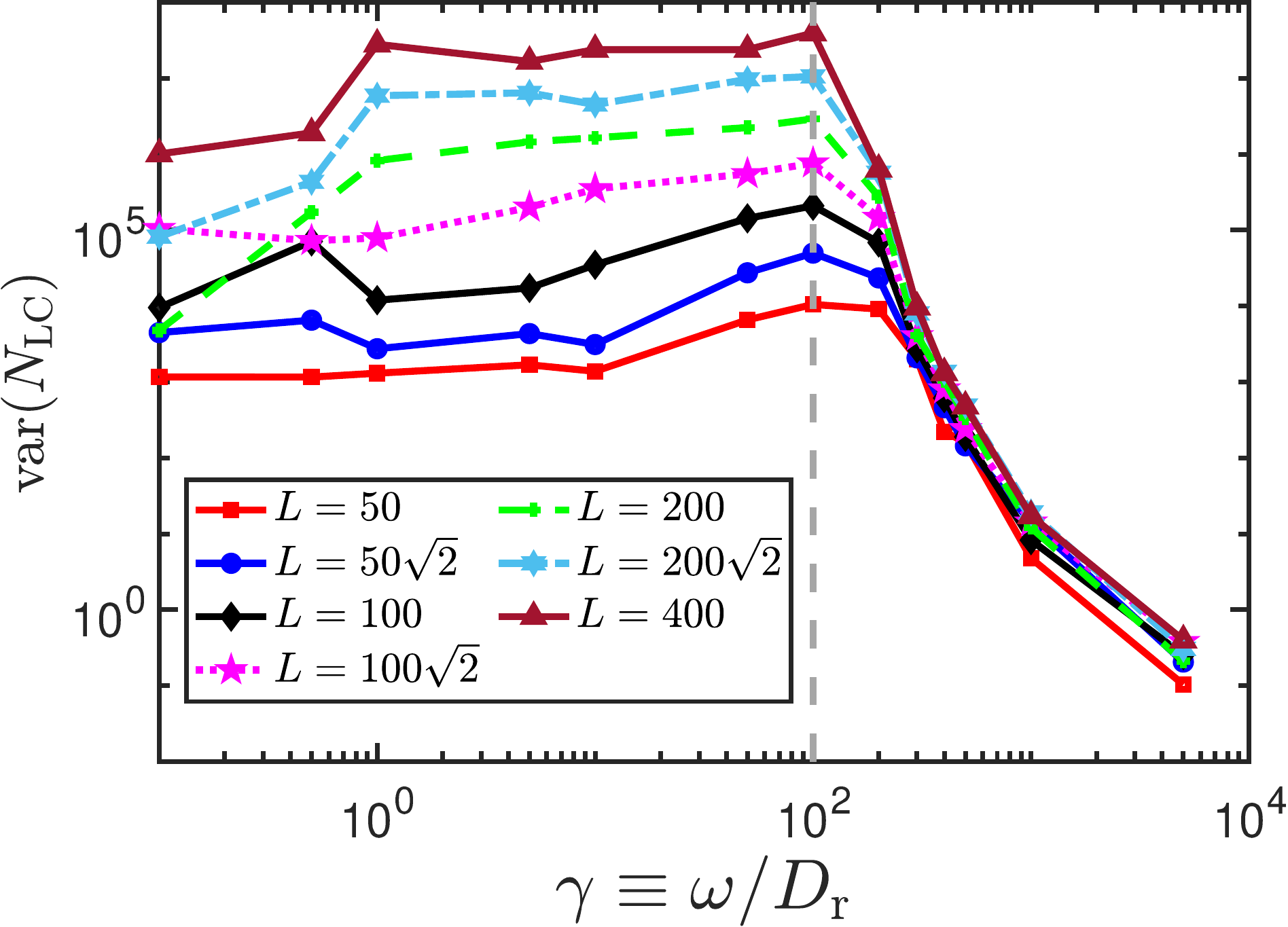}\\
(b)\\
\end{tabular}
\end{center}
\caption{(a) Mean and (b) variance in the size of the largest cluster for a collection of RS disks. The number of disks in the periodic box, for the smallest to largest values of $L$, are $N=400,800,1600,3200,6400,12800,25600$. The diameter of the disks considered in the figure is $d=2.0$, with $U_{0}=0.1$ and $D_{\text{r}}=2\times 10^{-4}$.}
\label{fig:length_dep_clust_stats}
\end{figure}

{The clustering dynamics of conventional ABPs is affected by the size of the simulation box~\cite{Patch2017,Patch2018}, particularly when the persistence length of the disk $\ell$ given by
\begin{equation}\label{eq:const_pers}
\ell=U_{0}/D_{r}
\end{equation}
is small compared to $L$. We have performed simulations for several values of the box length, at a constant area fraction of $\phi=0.5$. We have performed these computations using HOOMD-blue on disks with diameter $d=2.0$. This is in contrast to disks with $d=0.4$ dimensionless units used in studies of systems with $N\leq1600$. Additionally, we use $D_{r}=2\times10^{-4}$ for the disks with $d=2.0$ to maintain the P\'{e}clet number at $Pe=3U_{0}/dD_{\mathrm{r}}=750$. As illustrated in Figs.~\ref{fig:length_dep_clust_stats}~(a) and (b), the mean and variance in the size of the largest cluster is clearly affected by the simulation box length. Finite size effects are most strongly perceived in the low-$\gamma$ region in Fig.~\ref{fig:length_dep_clust_stats}~(b), where the persistence length should be close to that given by eq.~(\ref{eq:const_pers}), since the frequency of speed reversals is smaller than the rotational diffusion coefficient. At larger values of $\gamma$, however, for which the disks undergo several reversals in direction and speed before changing directions due to orientational diffusion, the persistence length is expected to be smaller than eq.~(\ref{eq:const_pers}), and the effect of the box length should consequently be much less pronounced. That is, for $\gamma>1$, one would replace the rotational diffusion coefficient with the frequency $\omega$ in the definition of the persistence length. Indeed, a universal feature for all the datasets in Fig.~\ref{fig:length_dep_clust_stats}~(b), is the appearance of a peak in the variance at $\gamma=100$.  We therefore believe that the observation of phase transition in a collection of RS disks at $\gamma=\mathcal{O}(100)$ is robust to finite-size effects.}

\begin{figure*}[t]
\begin{center}
\begin{tabular}{c c c}
\includegraphics[width=2.3in,height=!]{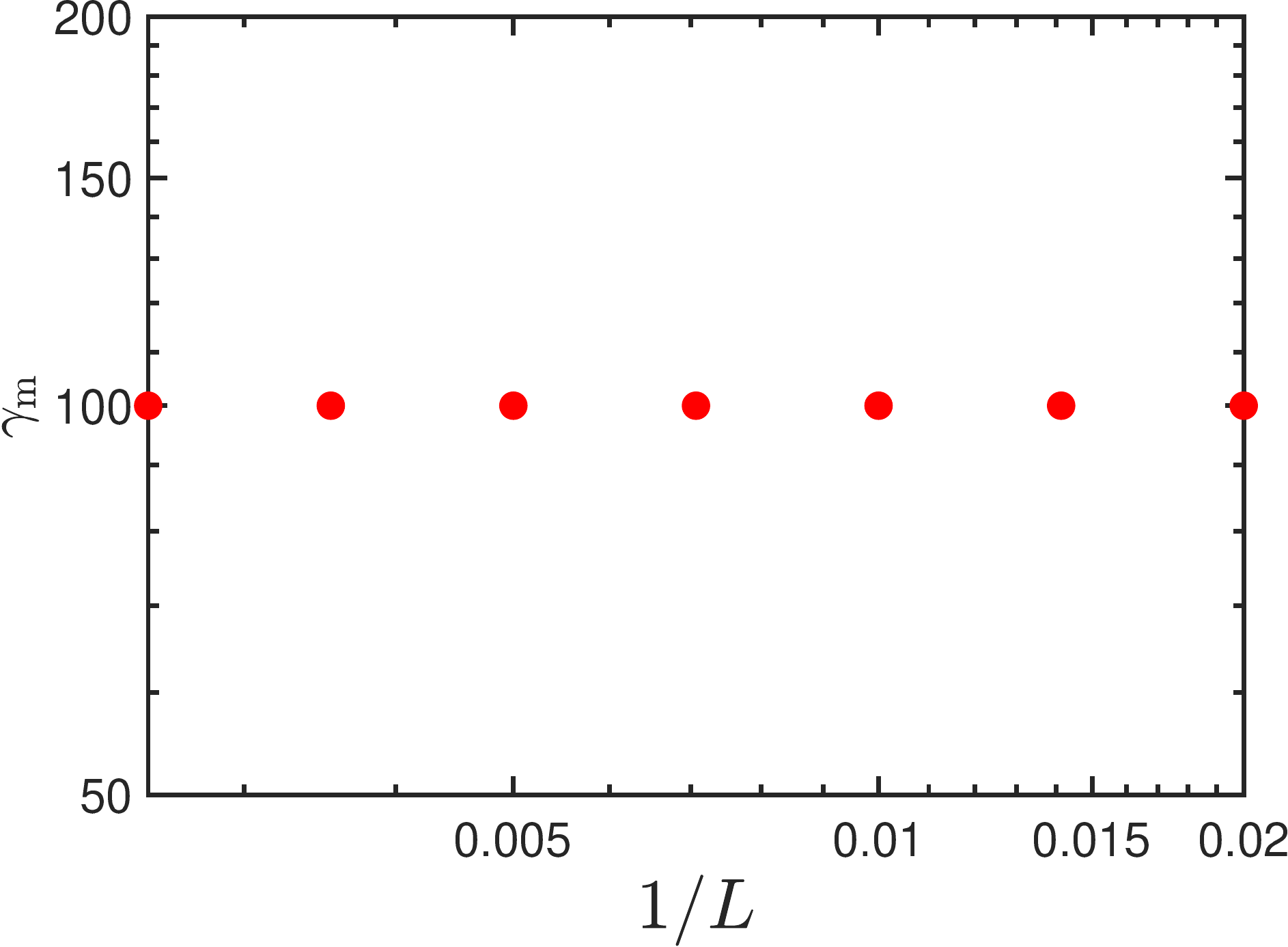}&
\includegraphics[width=2.3in,height=!]{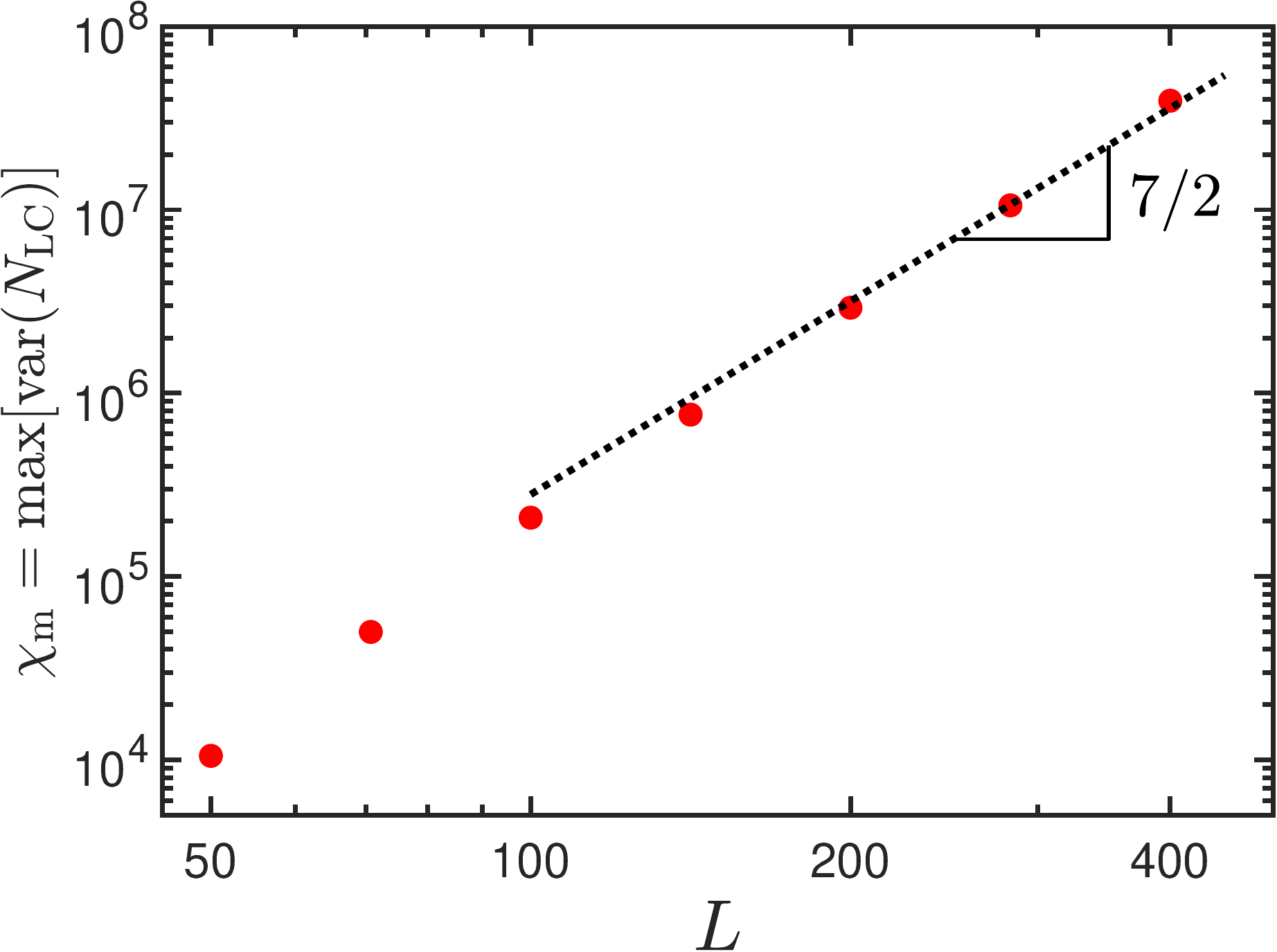}&
\includegraphics[width=2.3in,height=!]{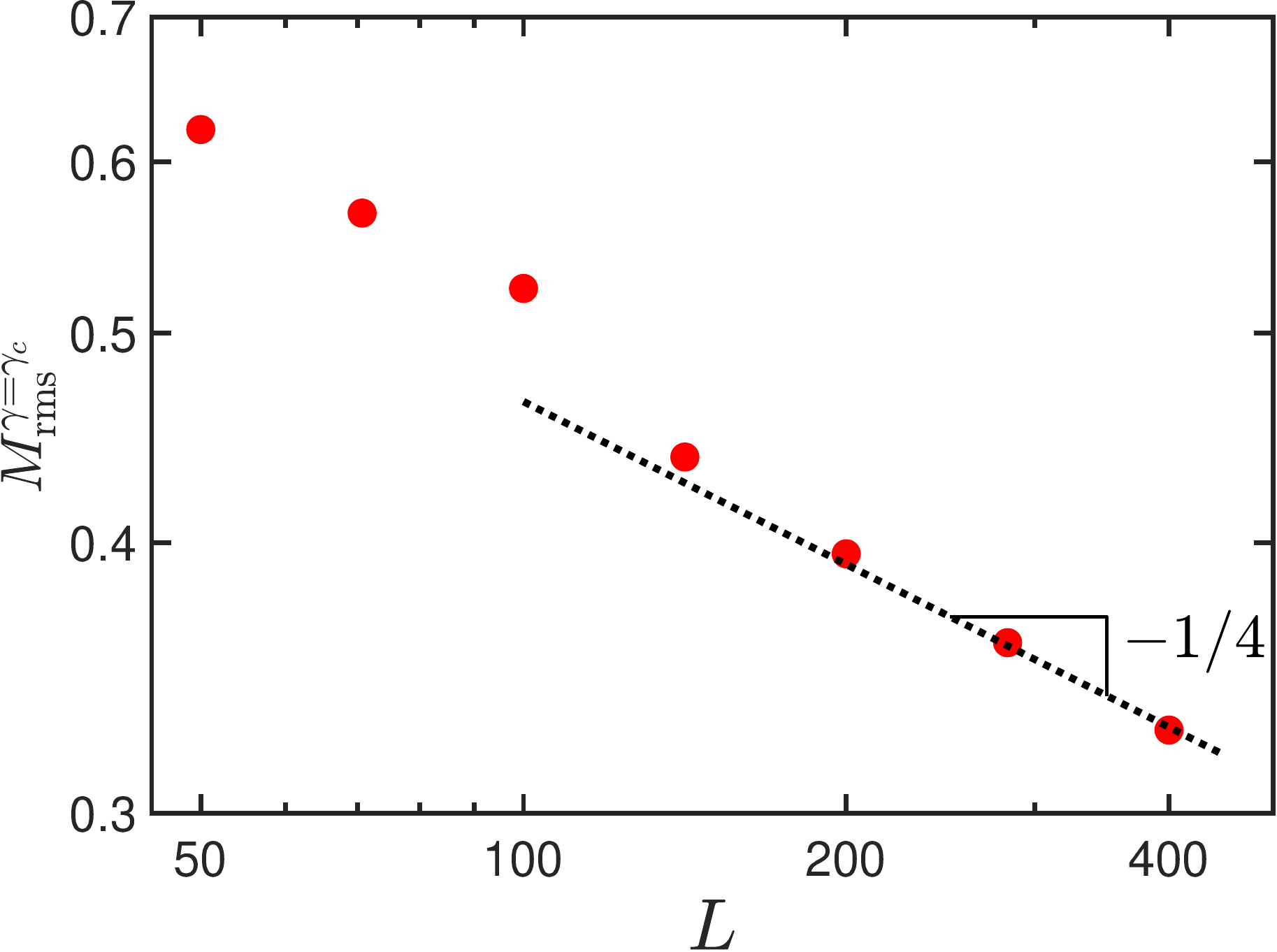}\\
(a)&(b)&(c)\\
\end{tabular}
\end{center}
\caption{Calculation of the critical exponents for a collection of RS disks: (a) Frequency at which the maximum in susceptibility is observed for a given box length, against the inverse of the box length, (b) peak susceptibility as a function of box length and (c) root-mean-square value of the order parameter measured at the critical frequency, as a function of box length. The data points are extracted from Fig.~\ref{fig:length_dep_clust_stats}.}
\label{fig:crit_exp_calc}
\end{figure*}

{We next comment upon the similarities {and differences} between the order-disorder transition in a collection of RS disks as the frequency is varied with that observed in the 2D Ising model as the temperature is varied~\cite{Landau1976,Chandler1987,DillBro2010}, by drawing an analogy between the disks in the gas phase (cluster) to spins in disordered (ordered) states.} {For a second-order phase transition, the following scaling laws hold~\cite{Landau1976,Landau1985,Binder1987} for the order parameter $M$, susceptibility $\chi$ and the correlation length $\xi$ in the vicinity of the critical temperature $T_{\text{c}}$:
\begin{align}
M&\sim|T-T_{\text{c}}|^{z_1}\\[5pt]
\chi&\sim|T-T_{\text{c}}|^{-z_2}\\[5pt]
\xi&\sim|T-T_{\text{c}}|^{-z_3}
\end{align}
The values of these exponents in the 2D Ising model~\cite{Binder1987,Cowan2021} are $z_{1}=1/8,\,z_{2}=7/4,\,z_{3}=1$. The calculation of the critical exponents from numerical simulations on finite-sized boxes is a well-covered topic in the literature~\cite{Landau1976,Landau1985,Binder1987}, so only a brief explanation is offered below~\cite{kari_2003}: the susceptibility is plotted as a function of the temperature for a range of lattice sizes ($L$), and the temperature $T_{\text{m}}$ at which the maximum in the susceptibility ($\chi_{\text{m}}$) occurs for a given lattice size is located. Using this information, it is possible to calculate $z_{2},z_{3}$ by fitting the following equations to the data points:
{\begin{equation}\label{eq:z_3_eq}
T_{\text{m}}=T_{\text{c}}-c_{3}L^{p};\,p=-1/z_{3};
\end{equation}
\begin{equation}\label{eq:z_2_eq}
\chi_{\text{m}}=c_{2}L^{z_2/z_3}
\end{equation}}
{For the calculation of $z_{1}$, one begins by defining the root-mean-square value of the order parameter as $M_{\text{rms}}=\sqrt{\left<M^2\right>}$, and plotting this value evaluated at the critical temperature for a range of box lengths, and fitting the following equation
\begin{equation}\label{eq:z_1_eq}
M^{T=T_{\text{c}}}_{\text{rms}}=c_{1}L^{-z_{1}/z_{3}}
\end{equation}}
The above procedure could be followed for a system of RS disks to extract the critical exponents, by appropriately replacing the temperature by the scaled frequency $\gamma$ in eqs.~(\ref{eq:z_3_eq})-(\ref{eq:z_1_eq}), and defining the order parameter $M=\left<N_{\text{LC}}\right>/N$ and susceptibility $\chi=N^2\left[\left<M^2\right>-\left<M\right>^2\right]=\text{var}(N_{\text{LC}})$. The estimation of critical exponents for a collection of RS disks is illustrated in Fig.~\ref{fig:crit_exp_calc}.}

{Unlike the Ising model in which the maximum in the susceptibility shifts with the size of the lattice, the peak in susceptibility for RS disks is observed at the same frequency, $\gamma_{\text{m}}$ for all the box lengths considered in the present work [Fig.~\ref{fig:crit_exp_calc}~(a)]. This precludes the calculation of $z_{3}$ in our system. It is possible to extract $z_1/z_3$ and $z_2/z_3$ as shown in Fig.~\ref{fig:crit_exp_calc}~(b) and (c), respectively. Due to the unavailability of $z_{3}$, we cannot conclude that the phase-transition in a system of RS disks is of second-order. Furthermore, the system does not belong to the same universality class as the 2D Ising model since the critical exponents are different.}

\section{\label{sec:prob_speed}Probabilistic speed evolution}

PM have derived analytically the expressions for the mean-squared displacement (MSD), $\left<\bm{r}^2(t)\right>$,  and the velocity autocorrelation (VAC) of an active disk with stochastic speed fluctuations, respectively, as follows
\begin{equation}\label{eq:msd_rel_pm}
\begin{split}
\left<\bm{r}^2(t)\right>&=\dfrac{2\left<v\right>^2}{D_{\text{r}}^2}\left[D_{\text{r}}t-1+e^{-D_{\text{r}}t}\right]\\
&+\dfrac{2\sigma^2}{\left(D_{\text{r}}+\beta\right)^2}\left[\left(D_{\text{r}}+\beta\right)t-1+e^{-\left(D_{\text{r}}+\beta\right)t}\right],
\end{split}
\end{equation}
\begin{equation}\label{eq:vac_rel_pm}
\begin{split}
\left<\bm{v}(t)\cdot\bm{v}(0)\right>=\left<v\right>^2 e^{-D_{\text{r}}t}+\sigma^2 e^{-\left(D_{\text{r}}+\beta\right)t}.
\end{split}
\end{equation}
The metric $\mu=\sigma/\left<U\right>$ characterizes the width of the stationary speed distribution, and $\mu=5.72$ for the power-law distribution considered in the present work. In terms of the scaled quantities $\mu$, $\gamma$, $\xi=rD_{\text{r}}/\left<U\right>$ and $\tau=D_{\text{r}}t$, we may recast eqs.~(\ref{eq:msd_rel_pm})-(\ref{eq:vac_rel_pm}) as
\begin{equation}\label{eq:msd_scaled_pm}
\left<\xi^2\right>=2\left[\tau-1+e^{-\tau}\right]+2\left(\dfrac{\mu}{1+\gamma}\right)^2\left[\gamma+e^{-(1+\gamma)\tau}\right],
\end{equation}
\begin{equation}\label{eq:vac_scaled_pm}
\dfrac{\left<\bm{v}(t)\cdot\bm{v}(0)\right>}{\left<v\right>^2}=e^{-\tau}+\mu^2e^{-\left(1+\gamma\right)\tau}.
\end{equation}
In the absence of speed fluctuations ($\mu=\sigma=0$), it is clear that eqs.~(\ref{eq:msd_scaled_pm})-(\ref{eq:vac_scaled_pm}) reduce to the familiar results for ABP~\cite{Zeitz2017,Ebbens2010}. 

\begin{figure}[t]
\begin{center}
\begin{tabular}{c}
\includegraphics[width=3in,height=!]{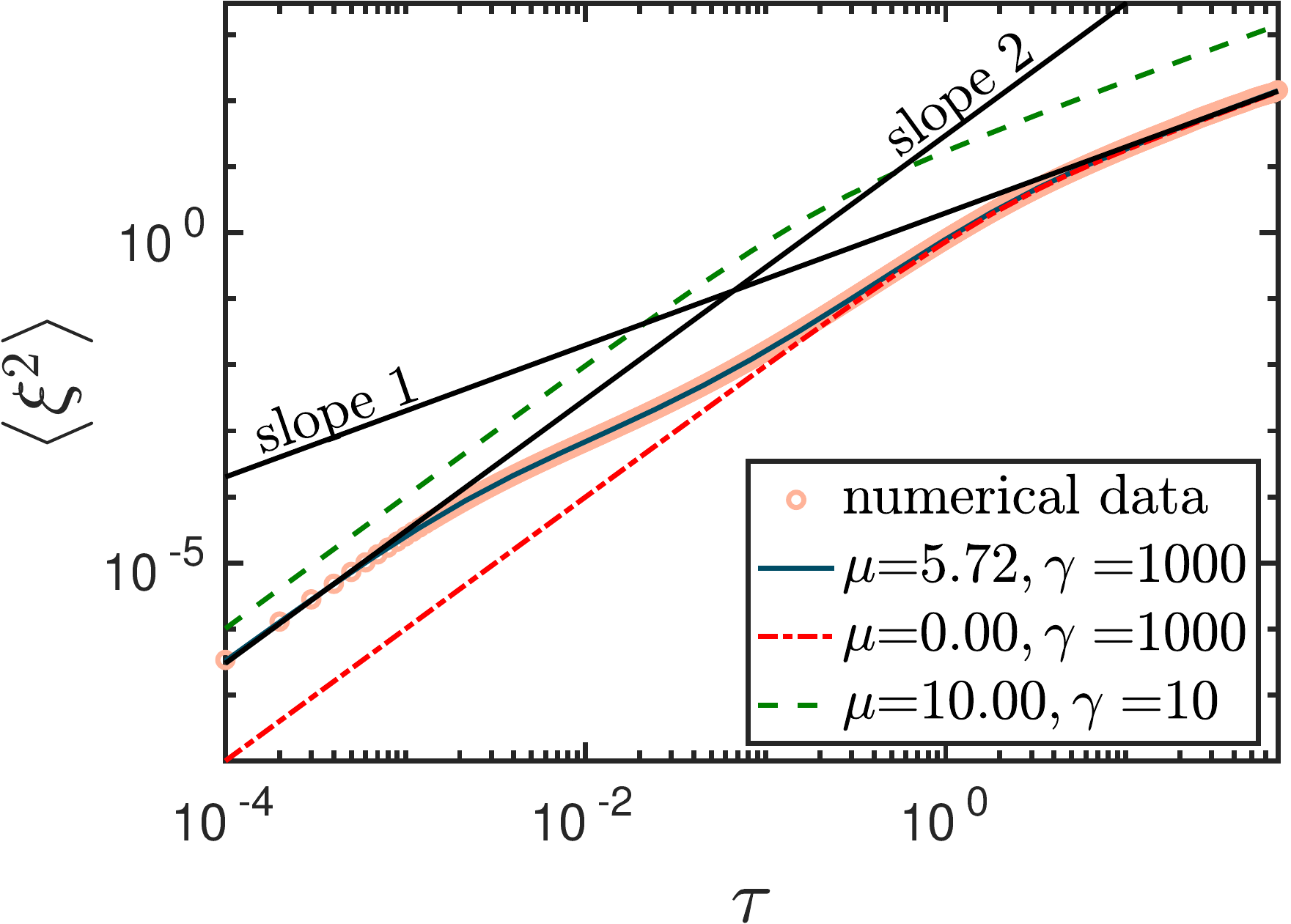}\\
(a)\\
\includegraphics[width=3in,height=!]{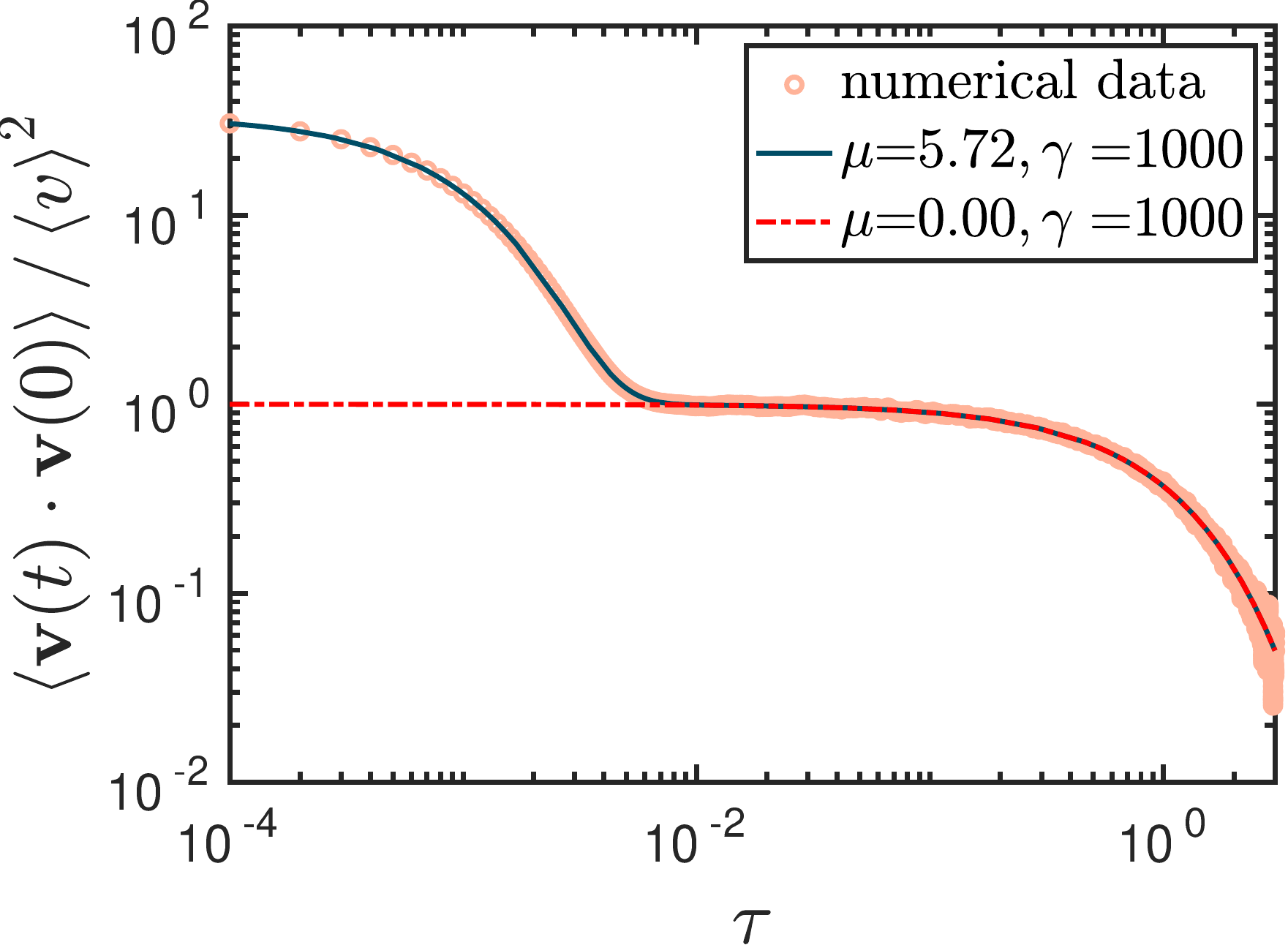}\\
(b)
\end{tabular}
\end{center}
\caption{(a) Mean square displacement and (b) velocity autocorrelation for a single PM disk with $U_{0}=0.1,n=10^4,\alpha=3/2,D_{\text{r}}=1.0,\beta=10^3$. The analytical results in (a) and (b) correspond to eqs.~(\ref{eq:msd_scaled_pm}) and~(\ref{eq:vac_scaled_pm}), respectively.}
\label{fig:pm_validn_single}
\end{figure}

\begin{figure}[t]
\begin{center}
\begin{tabular}{c}
\includegraphics[width=3in,height=!]{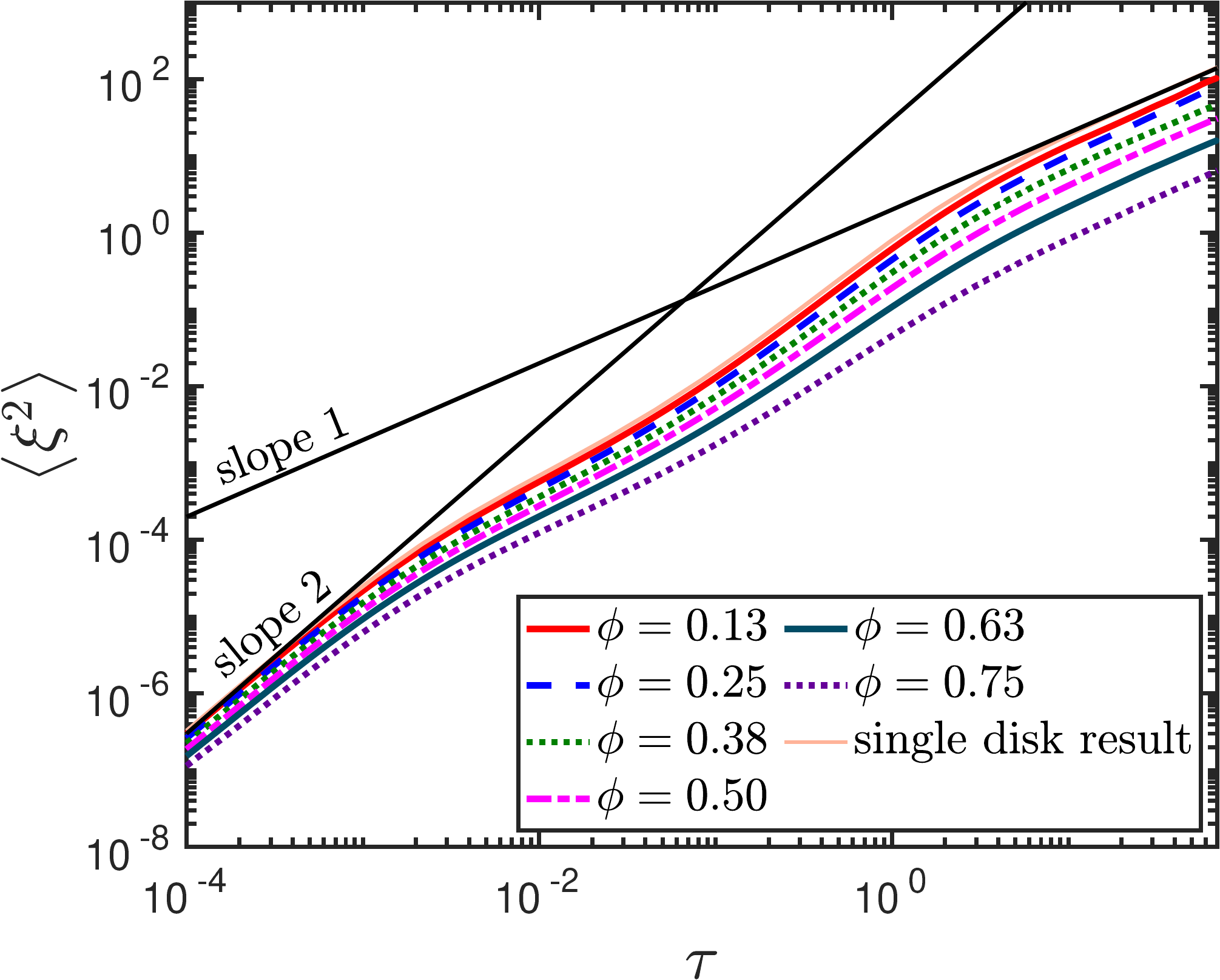}\\
(a)\\
\includegraphics[width=3in,height=!]{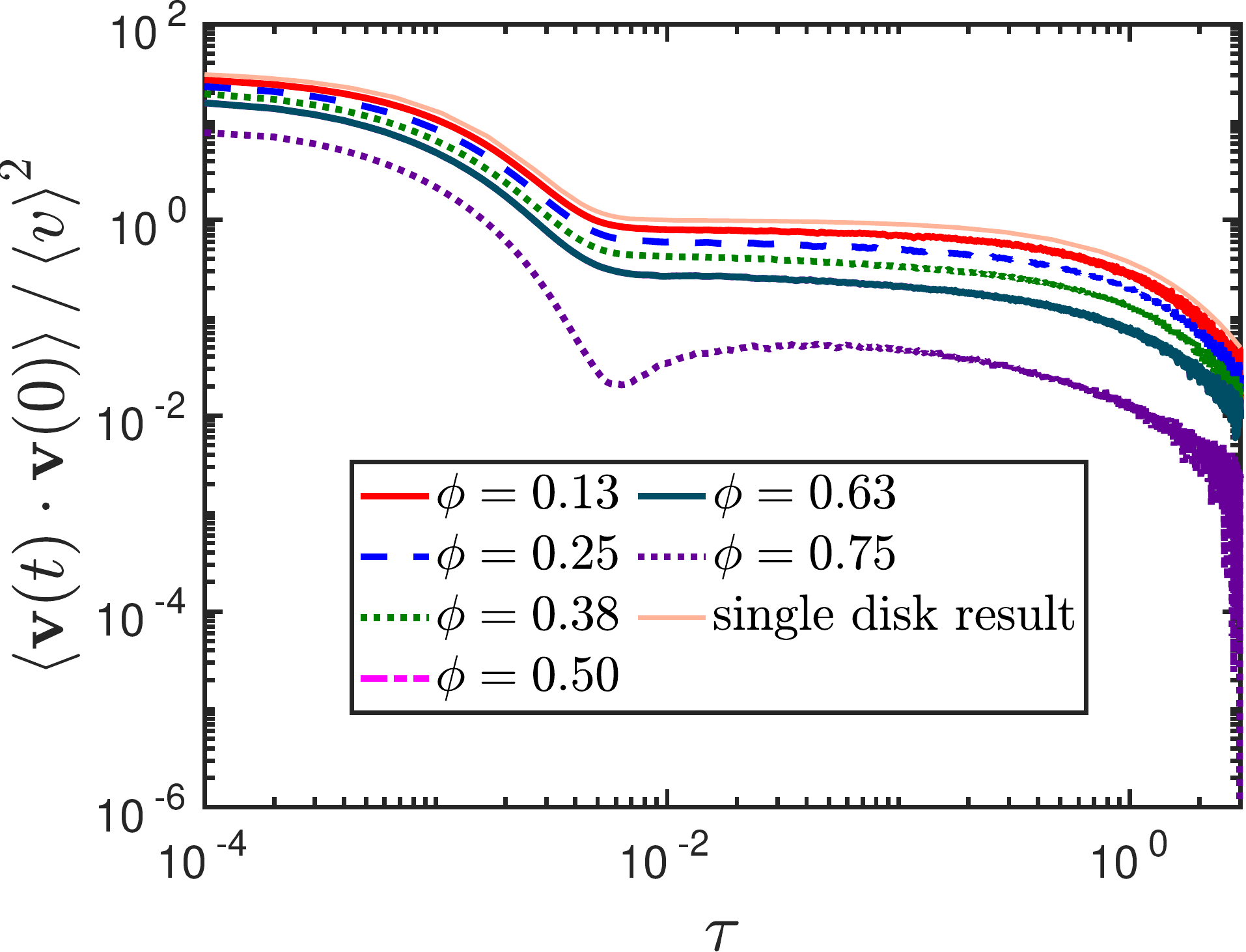}\\
(b)
\end{tabular}
\end{center}
\caption{(a) Mean square displacement and (b) velocity autocorrelation for a collection of PM disks at various values of the area fraction. The single disk results in (a) and (b) correspond to eqs.~(\ref{eq:msd_scaled_pm}) and~(\ref{eq:vac_scaled_pm}), respectively. The parameters for the PM disks used in this figure are $U_{0}=0.1,n=10^4,\alpha=3/2,D_{\text{r}}=1.0,\beta=10^3$. The velocity autocorrelation at various $\phi$ values are normalized by the analytically known $\left<v\right>^2$ for a single PM disk (cf.~eq.~\ref{eq:av_U_pwlaw})}
\label{fig:pm_area_frac}
\end{figure}

In fig.~\ref{fig:pm_validn_single}, our numerically simulated MSD and VAC are compared against these analytical results, and the agreement is excellent. The effects of speed fluctuations are most evident at short time-scales $\tau\ll1$. The long-time diffusivity $D_{0}$ of the PM disk is defined as follows, 
\begin{equation}
\begin{split}
D^{\text{PM}}_{0}=\dfrac{1}{2}\left[\dfrac{\left<v\right>^2}{D_{\text{r}}}+\dfrac{\sigma^2}{D_{\text{r}}+\beta}\right]=\dfrac{\left<v\right>^2}{2D_{\text{r}}}\left[1+\dfrac{\mu^2}{1+\gamma}\right].
\end{split}
\end{equation}
The increase in long-time diffusivity due to speed fluctuations is most clearly observed for systems with low values of $\gamma$. Physically, this corresponds to a disk being able to sample many speed fluctuations in the time scale for an orientation fluctuation. The diffusion constant for a single PM disk is bounded from below by the classical ABP result, which serves as the upper bound for the RS disks discussed previously. A detailed discussion of the various scaling regimes observed in the MSD of a single PM disk is available in ref.~\citenum{Peruani2007}, which is clearly much richer than for an ABP. The velocity autocorrelation is biexponential, with separate timescales associated with the relaxation of the speed and the orientation, as seen from eq.~(\ref{eq:vac_scaled_pm}).

\begin{figure}[t]
\centering
\includegraphics[width=3.1in,height=!]{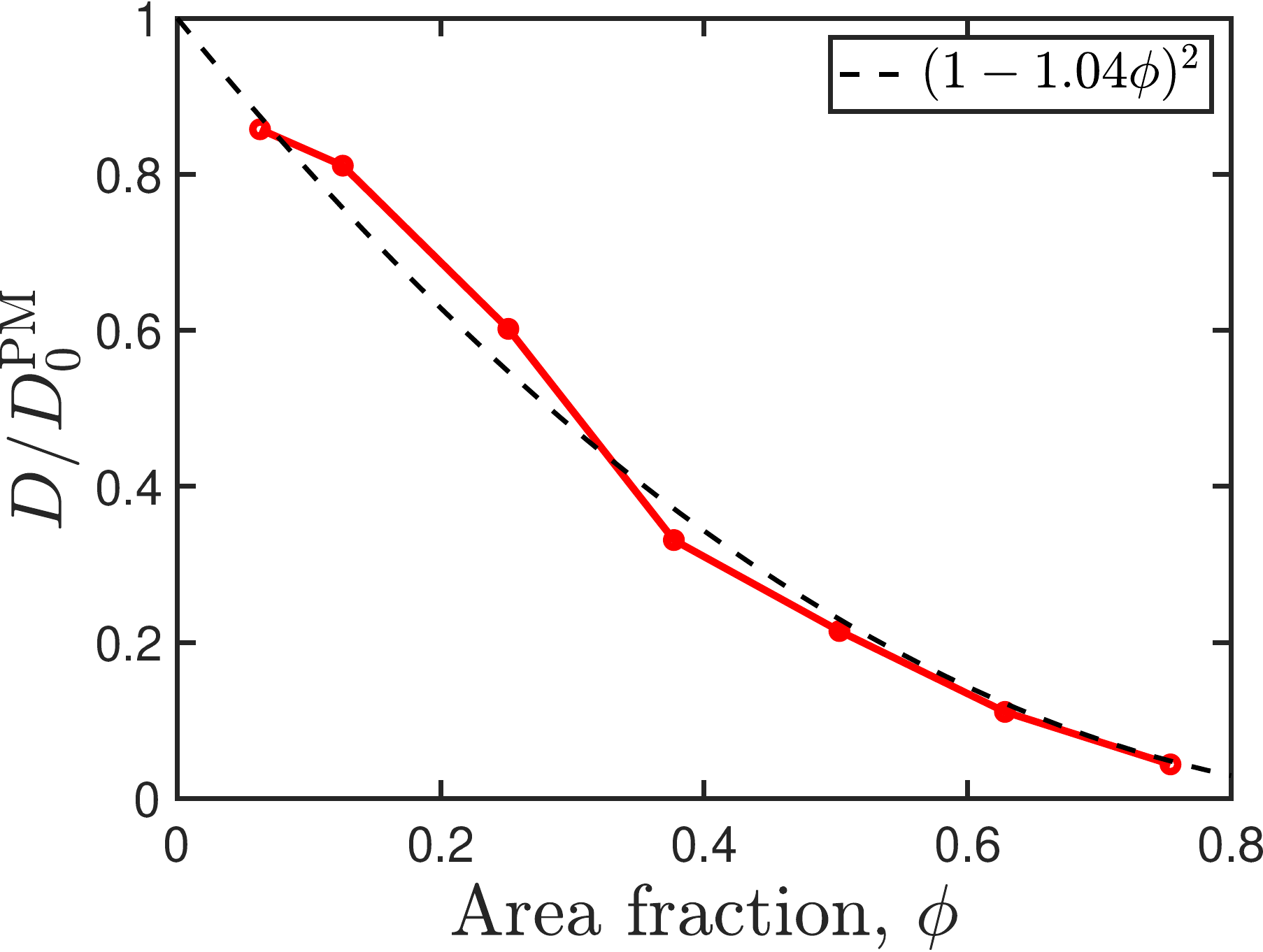}
\caption{Long-time diffusion constant for a collection of PM disks as a function of their area fraction. {The simulation parameters are identical to that used in Fig.~\ref{fig:pm_area_frac}.}}
\label{fig:diff_phi}
\end{figure}

In Fig.~\ref{fig:pm_area_frac}, the MSD and VAC of a collection of PM disks is plotted at various values of the area fraction. At lower values of the area fraction ($\phi<0.3$), there are two ballistic regimes in the MSD: the first is observed for $\tau\ll(1+\gamma)^{-1}$, while the second ballistic regime precedes the long-time asymptotic diffusive regime observed for $\tau\gg1$. With an increase in area fraction, however, the latter ballistic regime is seen to vanish. The transition between the two timescales in the VAC plot is smooth at low values of $\phi$. At the highest area fraction considered in Fig.~\ref{fig:pm_area_frac}~(b), however, the transition proceeds through a pronounced dip.

In Fig.~\ref{fig:diff_phi}, the long-time diffusion constant $D$ evaluated from the curves in Fig.~\ref{fig:pm_area_frac}~(a), is normalized by the single particle diffusivity $D^{\text{PM}}_{0}$ and plotted as a function of the area fraction. It is observed that the decrease in the scaled diffusivity with increasing area fraction can be reasonably approximated by the functional form $\left(1-b\phi\right)^2$, with the fitting constant given by $b\approx1.04$. A similar scaling has been reported for conventional ABPs~\cite{Stenhammar2013,Stenhammar2014}.

\begin{figure}
\centering
\includegraphics[width=3.1in,height=!]{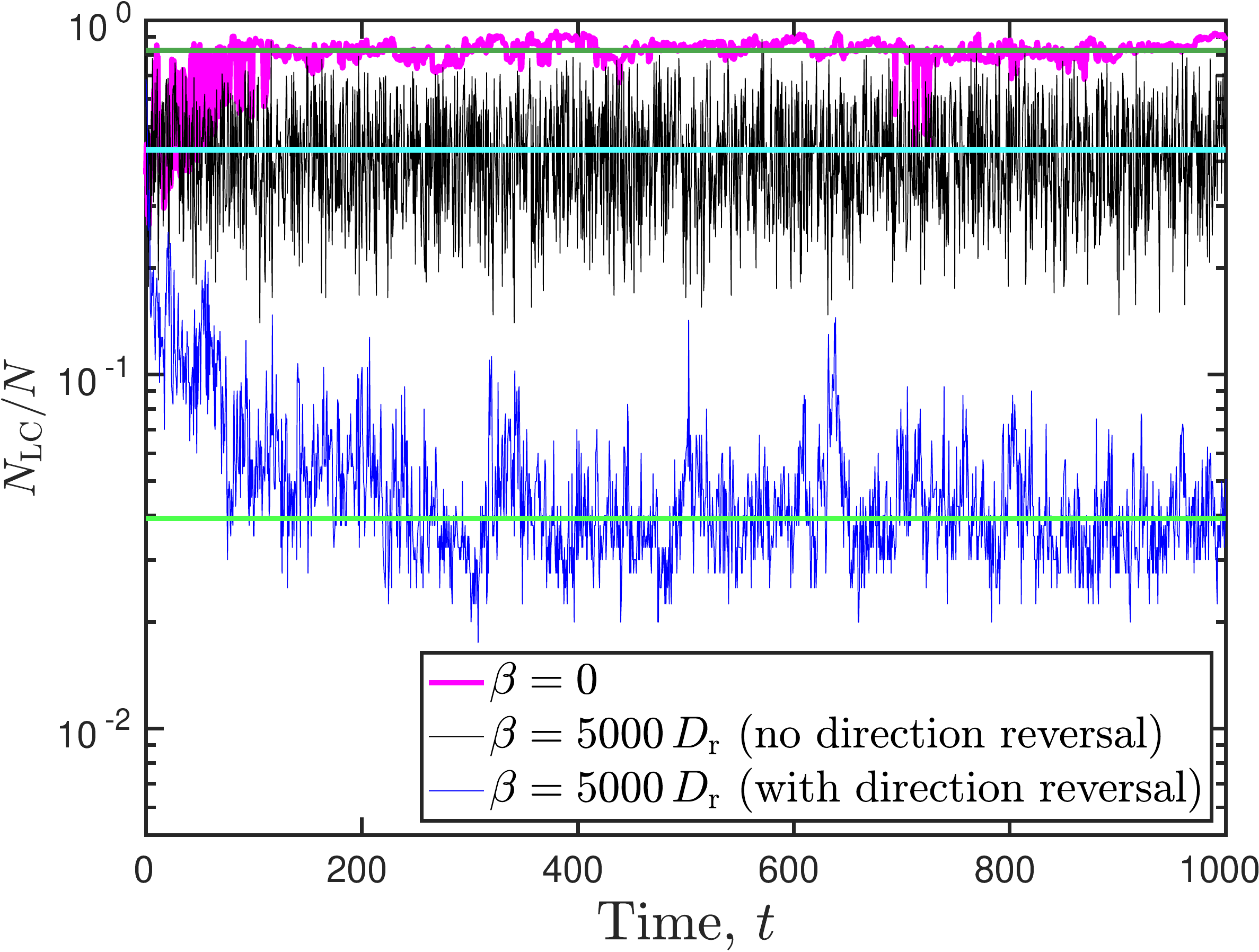}
\caption{Time series of the size of the largest cluster, for a collection of $N=400$ PM disks in a periodic box of $L=10$. The speeds for PM disks with direction reversal are drawn from a uniform distribution in the interval $[-0.1,0.1]$, while that for the case without direction reversal are drawn from a power-law distribution $P(U)\sim U^{-3/2}$ in the interval $[0.1,1000]$. The rotational diffusion constant is $D_{\text{r}}=10^{-3}$ for all the datasets. Horizontal lines indicate the mean size of the cluster.}
\label{fig:pm_eff_dir_rev}
\end{figure}

The effect of stochastic speed fluctuations on MIPS is examined next, for the two cases of speed-updates: with and without directional reversals. To simulate a reversal in the self-propulsion direction, the speeds of PM disks are sampled from a uniform distribution in the interval $[-0.1,0.1]$. Speeds picked from the power-law distribution discussed above are always non-negative and therefore correspond to the case without directional reversal. As seen from Fig.~\ref{fig:pm_eff_dir_rev}, while speed fluctuations without directional reversals cause a marginal decrease in the average size of the largest cluster (also see Supplementary Video 6), a speed-update protocol with directional reversals drives the system more strongly into the homogeneous state (also see Supplementary Video 8).

Lastly, in fig.~\ref{fig:comm_gamma_plot}, the cluster statistics for both PM and RS disks with directional reversals is plotted as a function of the dimensionless switching frequency $\gamma$. While the absolute values of the average cluster size and its variance are lower for PM disks, the variations of these statistics with respect to the switching rate are similar irrespective of the nature of the speed update protocol. The peak in the variance is also observed to occur at the same value of the switching frequency. At small values of $\gamma$, the fluctuations in the self-propulsion direction are slower than changes in the orientation, and thus regular MIPS is observed. However at sufficiently large values of $\gamma$, the disks undergo directional reversals faster than orientational fluctuations, and MIPS is suppressed. While one might suspect that an order-disorder transition would occur around $\gamma\sim\mathcal{O}(1)$, this is clearly not the case as seen from Fig.~\ref{fig:comm_gamma_plot}. The identification of the appropriate order parameter for this transition, which may involve a combination of $\gamma$, the area fraction and the self-propulsion speed of the individual disks, is reserved for future work.

\begin{figure}[t]
\begin{center}
\begin{tabular}{c}
\includegraphics[width=3in,height=!]{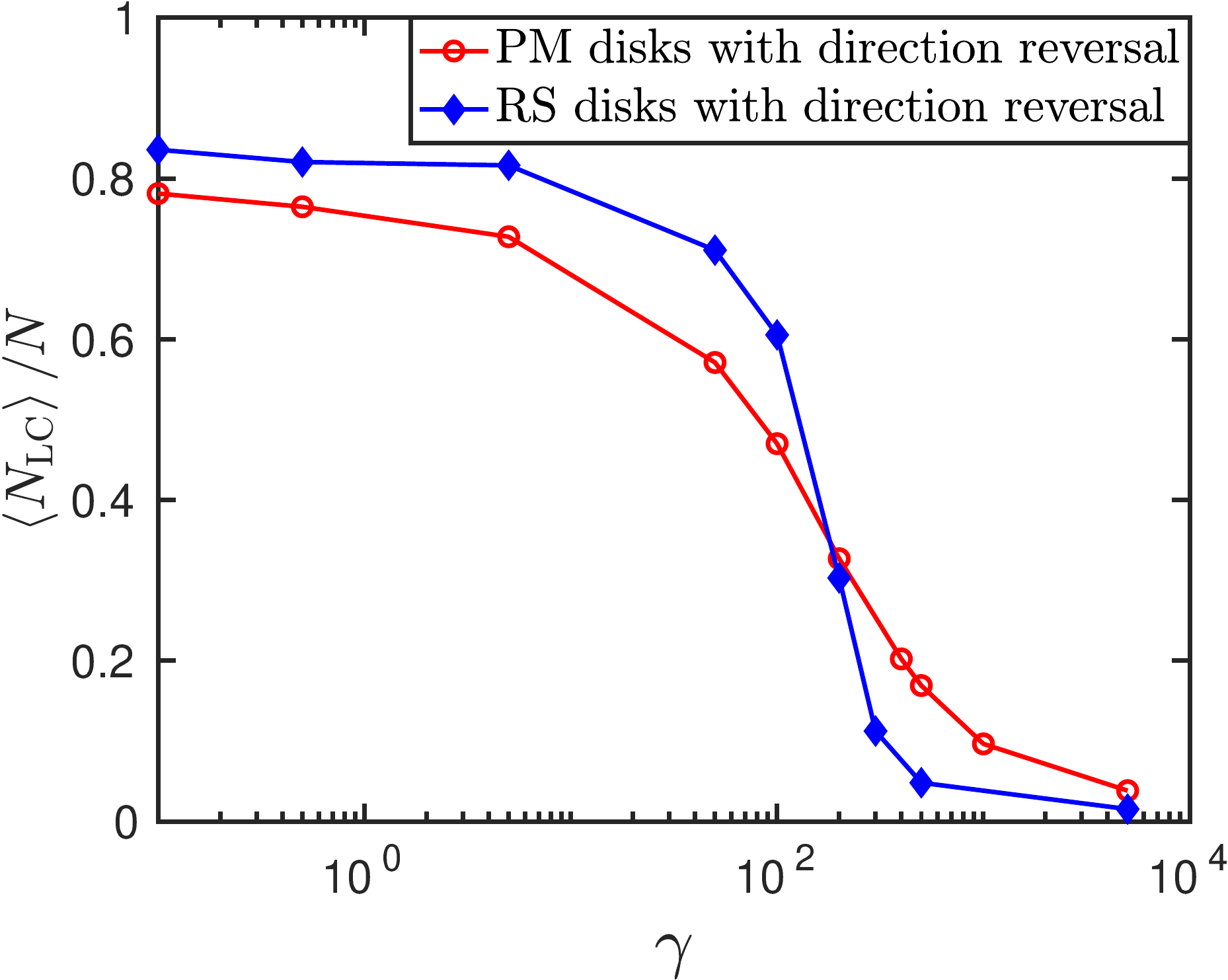}\\
(a)\\
\includegraphics[width=3in,height=!]{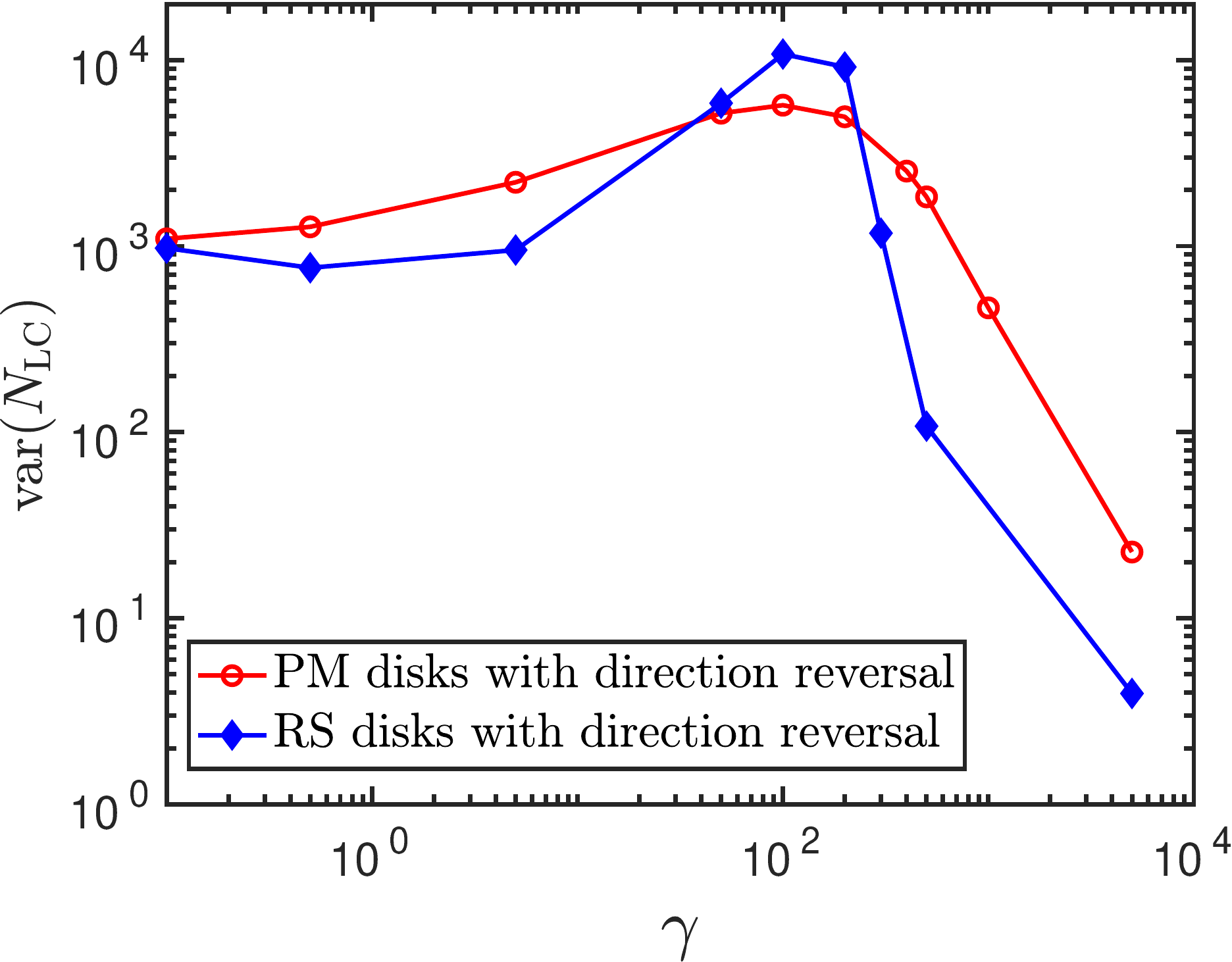}\\
(b)
\end{tabular}
\end{center}
\caption{(a) Mean and (b) standard-deviation of the steady-state cluster size for PM and RS disks undergoing directional reversal. A collection of $N=400$ disks in a periodic box length $L=10$ is considered. {Data for PM disks with direction reversal has been averaged over five independent realizations.}}
\label{fig:comm_gamma_plot}
\end{figure}

\section{\label{sec:concl}Conclusions}

Using a minimal model for an active particle that involves an update rule for the speed and orientation, we have studied the effects of speed fluctuations and direction reversals on the MIPS of a collection of such disks. For the case of a deterministic speed-fluctuation protocol, which involves a reversal in the swimming direction of the disks, there is a transition from a clustered to homogeneous phase as the frequency of speed variation is increased. {Remarkably, MIPS is observed for such disks at low frequencies even though the mean motility (in the time-averaged sense) is zero. The existence of the phase-transition is established through simulations of systems of various box lengths, and the critical exponents estimated from finite-size scaling. An analytical derivation for the critical frequency, and studying the effect of area fraction on the collective dynamics of disks with fluctuating swim-speed are potentially interesting questions for future work.} In the absence of reversals in the swimming direction, however, the order-disorder transition is absent. The feature of directional reversals suppressing phase-separation also holds for the case of PM disks undergoing stochastic fluctuations in speed. That is, we find that, irrespective of the precise nature of the speed-update protocol (deterministic or stochastic), the inclusion of directional reversals precipitates an order-disorder phase transition.   

{MIPS has also been observed for a collection of chiral active particles~\cite{DelJunco2018,Kreienkamp2022,Ma2022,Lei2023} whose time-rate of change of orientation has a deterministic and a diffusive component. Increasing the deterministic frequency of rotation of an individual agent is shown to cause a suppression of MIPS~\cite{Ma2022}. This similarity can be understood by recognizing that MIPS is sensitive to the effective persistence length of the active disk, which may be affected by changes in either the speed fluctuation frequency or the rotational frequency of the active agent. This explains the similarities in the frequency-dependent clustering observed in the present work and in a collection of chiral active particles.}

\begin{figure}
\begin{center}
\begin{tabular}{c}
\includegraphics[width=3in,height=!]{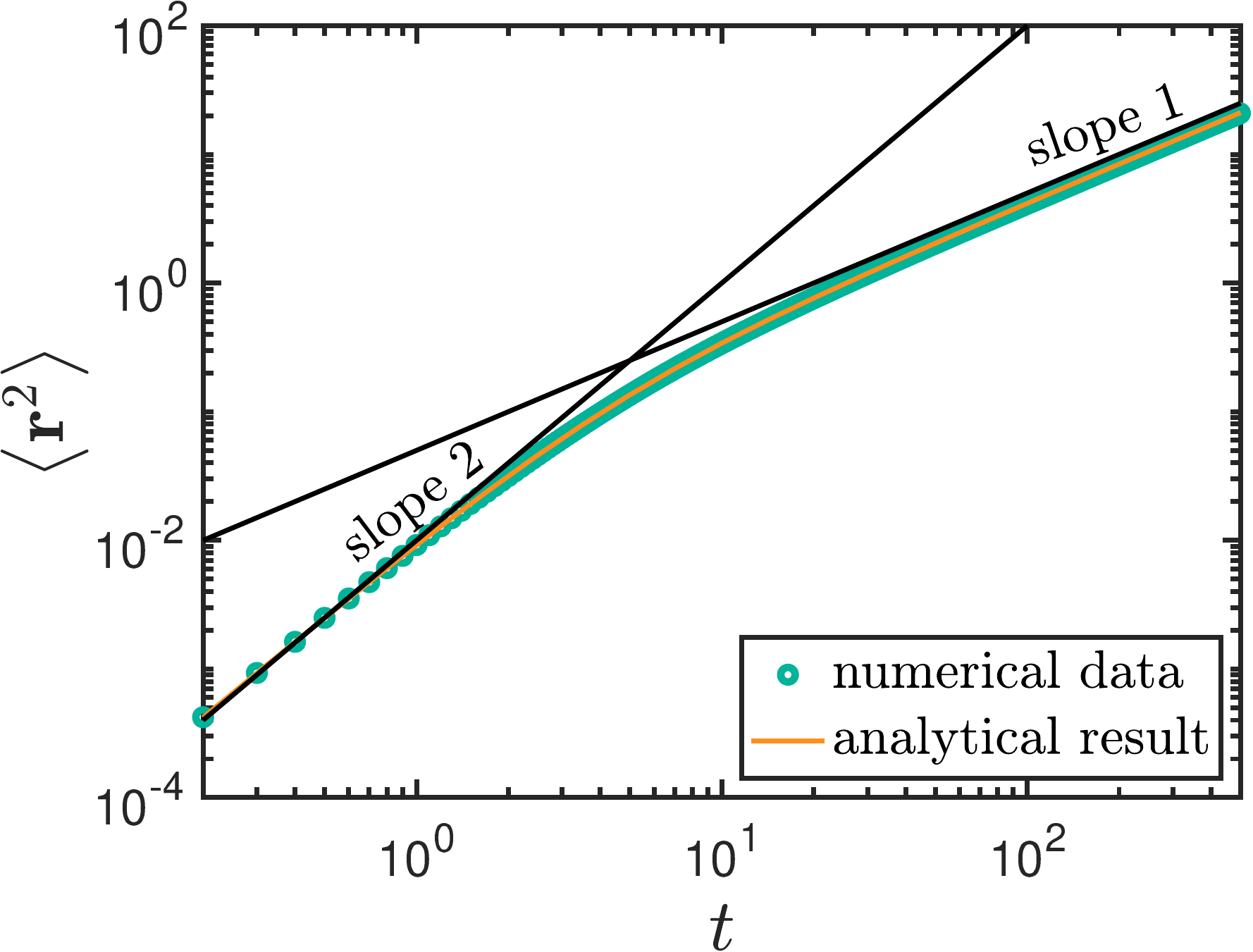}\\
(a)\\
\includegraphics[width=3in,height=!]{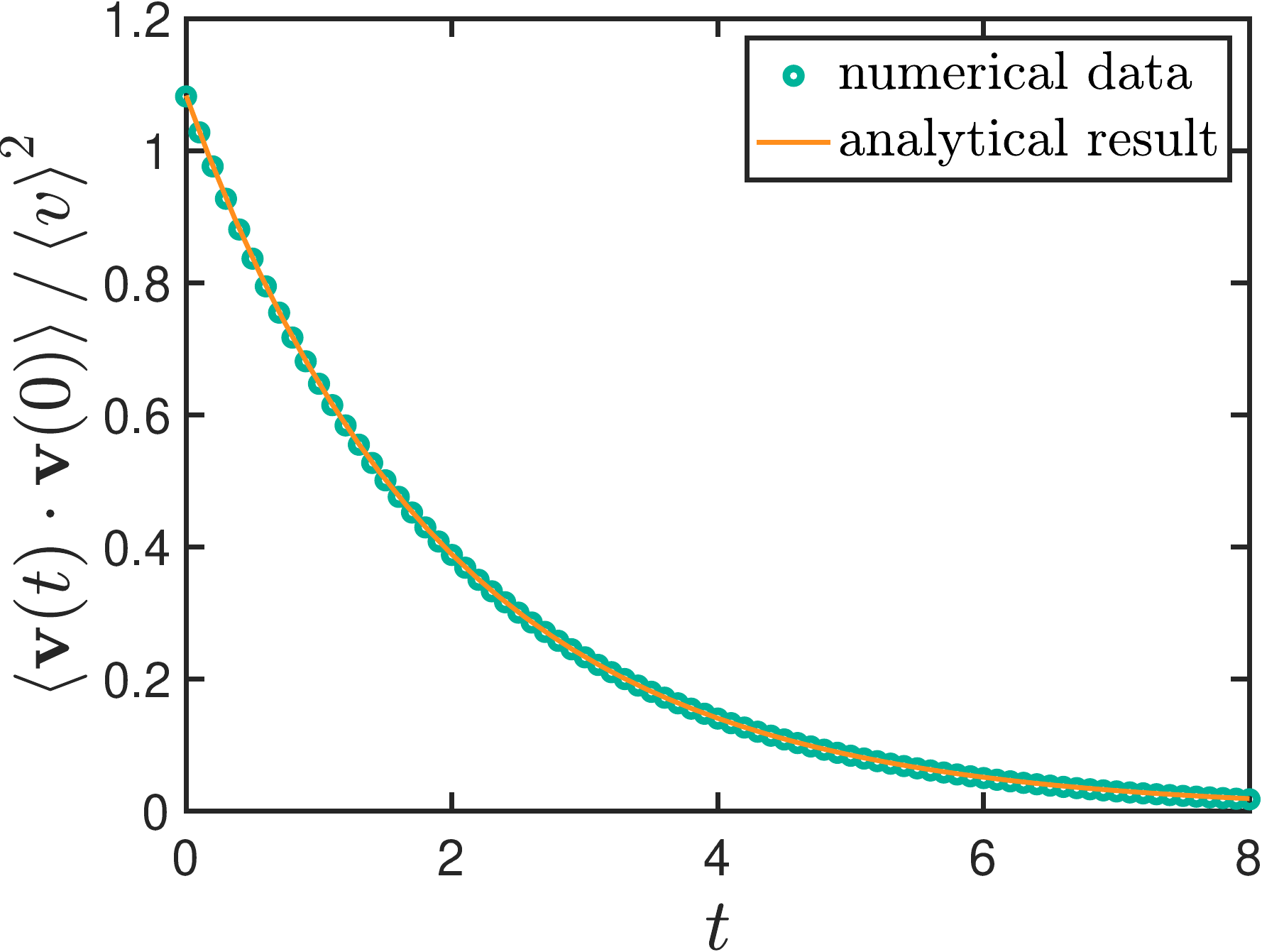}\\
(b)
\end{tabular}
\end{center}
\caption{(a) Mean square displacement and (b) velocity autocorrelation for a single PM disk, averaged over 300 independent realizations. The parameters for the PM disk used in this figure are $U_{0}=0.1,\delta=0.05,D_{\text{r}}=0.5, \beta=0.2$. The lines in (a) and (b) correspond to the analytical result given by eqs.~(\ref{eq:msd_rel_pm}) and~(\ref{eq:vac_rel_pm}), respectively. The numerical data in (b) is normalized by $\left<v\right>^2$ for a single PM disk (cf.~\ref{eq:mom_unif}). A timestep width of $\Delta t=0.1$ has been used for the simulation of trajectories associated with this plot.}
\label{fig:pm_validn_single_unif}
\end{figure}

{\section*{Author contributions}}
{R.K. and A.S.K designed research, performed research, contributed new analytic tools, analyzed data, and wrote the paper.}

\section*{Conflicts of interest}
There are no conflicts to declare.

\section*{Data and Code availability}
The simulations of PM and RS disks have been performed using custom-written MATLAB code, available freely on GitHub~\cite{Kailasham_PM_and_RS_2023}. {Larger scale simulations of RS disks were performed using a custom-written code~\cite{Kailasham_RS_HOOMD_2023} in HOOMD-blue}. The calculation of MSD and VAC was performed using a Fourier component analysis~\cite{Calandrini2011}, and the Python codes are available freely on GitHub~\cite{Kach_msd_vac_calc_2d_fft_2023}. Visualization of simulation data for the generation of videos in the Supplementary Material was performed using OVITO~\cite{Ovito2010}. Simulation data are available upon request from the authors.

\section*{Appendix}

\appendix

\section{\label{sec:unif_pm} PM disks with speeds drawn from a uniform distribution}

In this section, we consider PM disks whose speeds obey alternative stationary distribution compared to the one discussed in the main text. The speeds are drawn from a uniform random distribution in the interval $[U_{0}-\delta,U_{0}+\delta]$ with $\left(U_{0}-\delta\right)\geq0$. The moments of the distribution may be written as follows, 
\begin{equation}\label{eq:mom_unif}
\begin{split}
\left<v\right>=U_{0};\quad\sigma^2=\delta^2/3,
\end{split}
\end{equation}
giving
\begin{equation}
\mu=\dfrac{\delta}{\sqrt{3}U_{0}}.
\end{equation}
We consider disks with $U_{0}=0.1$ and $\delta=0.05$, which represents an allowable fluctuation of $50\%$ about the mean speed. However, these translate to a value of $\mu\approx0.3$, resulting in MSD and VAC curves that are nearly identical to the standard ABP result, as seen from fig.~\ref{fig:pm_validn_single_unif}. Upon comparison of figs~\ref{fig:pm_validn_single} and~\ref{fig:pm_validn_single_unif}, it is therefore evident that the variance of the stationary speed distribution must be sufficiently larger than its mean value to observe a significant deviation from the classical ABP result.

\section*{Acknowledgements}
We gratefully acknowledge the support of the Charles E. Kaufman Foundation of the Pittsburgh Foundation (Grant No. 1031373-438639).

\bibliography{ms_v2} 

\end{document}


\beginsupplement

\newcommand{\blueline}{\raisebox{2pt}{\tikz{\draw[-,blue,solid,line width = 1.5pt](0,0) -- (5mm,0);}}}

\newcommand{\blackline}{\raisebox{2pt}{\tikz{\draw[-,black,dashed,line width = 1.5pt](0,0) -- (5mm,0);}}}



\title{Supplementary Material for: Effect of speed fluctuations on the collective dynamics of active disks}


\author{R. Kailasham}
\affiliation{Department of Chemical Engineering, Carnegie Mellon University, Pittsburgh, PA 15213, USA}
\author{Aditya S. Khair}
\email{akhair@andrew.cmu.edu}
\affiliation{Department of Chemical Engineering, Carnegie Mellon University, Pittsburgh, PA 15213, USA}


\date{\today}


\maketitle


%
The main paper uses numerical simulations of reduced order models, to investigate the effect of speed fluctuations on the collective dynamics of active disks. Section~\ref{sec:vid_list} contains a description of the sample videos associated with the paper, and Section~\ref{sec:hmd_matlab} provides a comparison of the cluster statistics obtained using MATLAB and HOOMD simulation routines. 

\section{\label{sec:vid_list}List of videos}
All videos generated using OVITO~\cite{Ovito2010} and recorded for $N=400$ active disks in a periodic box of length $L=10$. The rotational diffusion constant is $D_{\text{r}}=10^{-3}$ and $U_{0}=0.1$ in all the cases. 
\begin{enumerate}
\item Video S1: Base case simulation indicating motility induced phase-separation in a suspension of active disks moving at a \textit{constant} self-propulsion speed of $U_{0}$. Video sped up by a factor of 10. 
\item Video S2: RS disks undergoing directional reversal in self-propulsion, i.e., $U(t)=U_{0}\cos(\omega t)$, with $\omega=5\,D_{\text{r}}$. Video sped up by a factor of 10.
\item Video S3: RS disks undergoing directional reversal in self-propulsion, i.e., $U(t)=U_{0}\cos(\omega t)$, with $\omega=500\,D_{\text{r}}$. Video sped up by a factor of 10.
\item Video S4: RS disks undergoing directional reversal in self-propulsion, i.e., $U(t)=U_{0}\cos(\omega t)$, with $\omega=5000\,D_{\text{r}}$. Video sped up by a factor of 10.
\item Video S5: RS disks without directional reversal in self-propulsion, i.e., $U(t)=|U_{0}\cos(\omega t)|$, with $\omega=5000\,D_{\text{r}}$. Video sped up by a factor of 100.
\item Video S6: PM disks without directional reversal, i.e.,  their speeds are drawn from a power-law distribution $P(U)\sim U^{-3/2}$ in the range $[0.1,1000]$, at update instances governed by a Poissonian process with rate constant $\beta=5000\,D_{\text{r}}$. Video sped up by a factor of 100.
\item Video S7: PM disks without directional reversal, i.e.,  their speeds are drawn from a power-law distribution $P(U)\sim U^{-3/2}$ in the range $[0.1,1000]$, at update instances governed by a Poissonian process with rate constant $\beta=40000\,D_{\text{r}}$. Video sped up by a factor of 100.
\item Video S8: PM disks with directional reversal, i.e.,  their speeds are drawn from a uniform distribution in the range $[-0.1,0.1]$, at update instances governed by a Poissonian process with rate constant $\beta=5000\,D_{\text{r}}$. Video sped up by a factor of 10.
\end{enumerate}

\section{\label{sec:hmd_matlab}Comparison between HOOMD-blue and MATLAB results}

\begin{figure*}[h]
\begin{center}
\begin{tabular}{c c}
\includegraphics[width=3in,height=!]{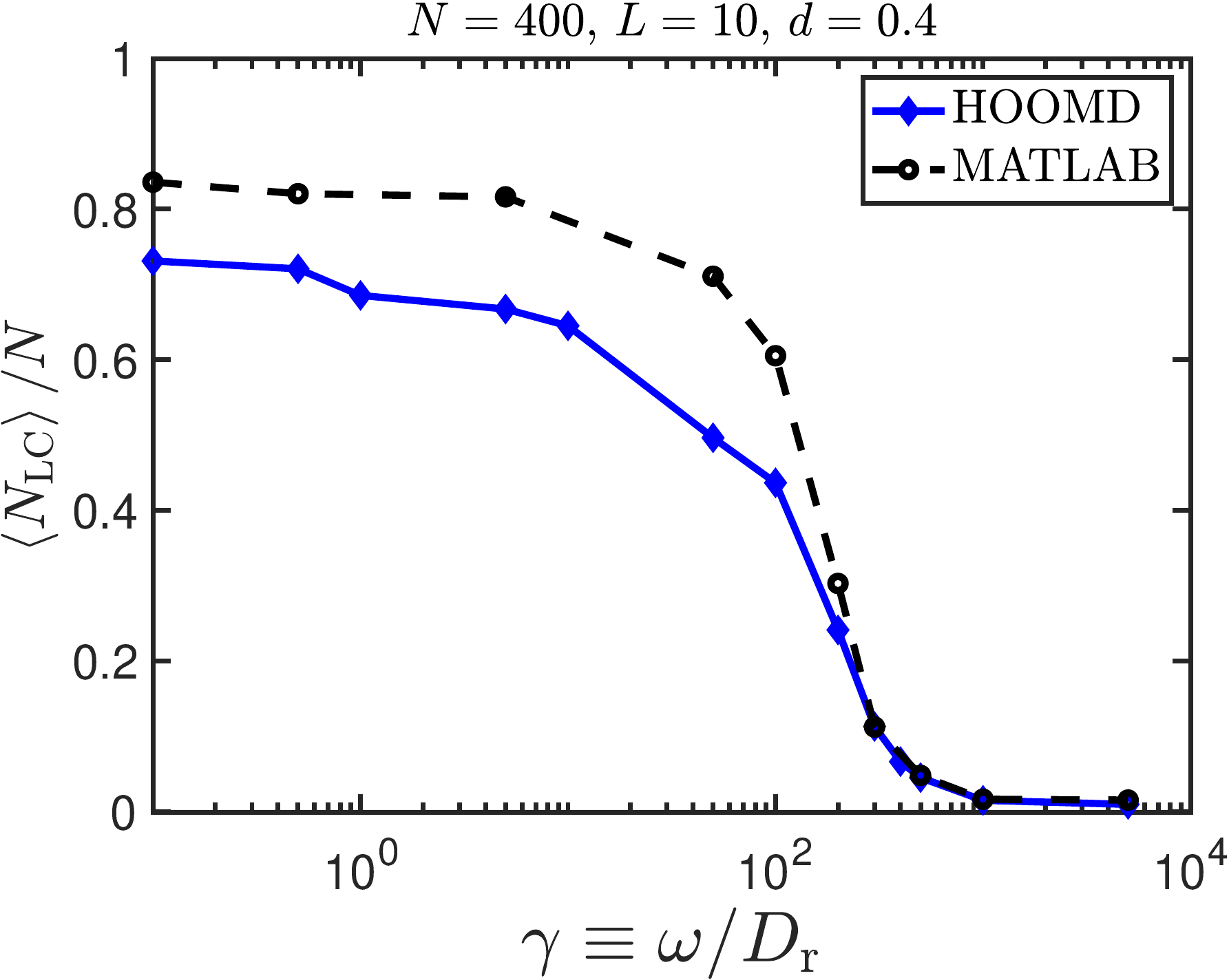}&
\includegraphics[width=3in,height=!]{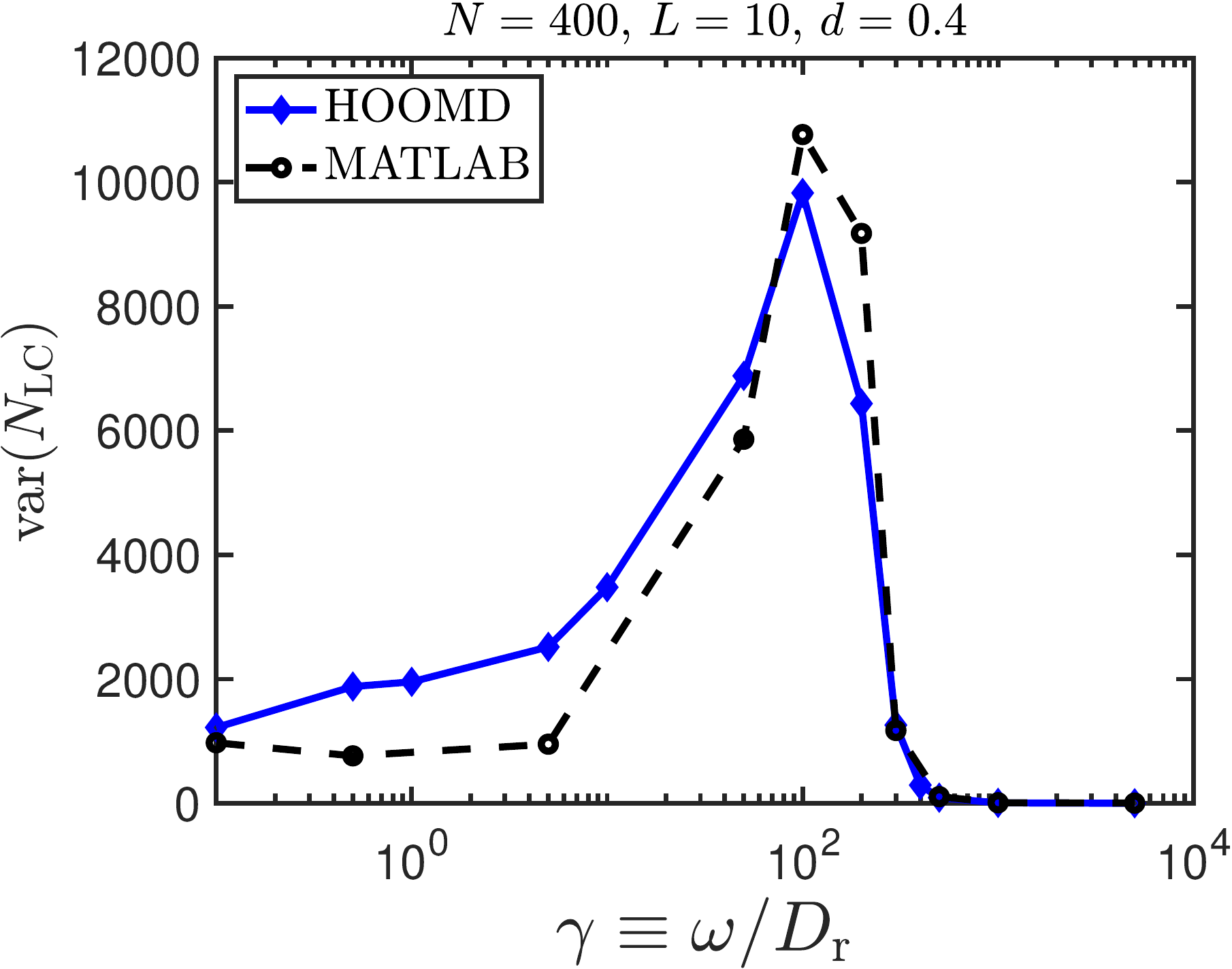}\\
(a)&(b)\\
\end{tabular}
\end{center}
\caption{(a) Time-averaged mean and (b) variance of cluster-size for a collection of RS disks simulated using HOOMD-blue and MATLAB.}
\label{fig:compare_stats}
\end{figure*}

The main paper contains simulation results for a collection of RS disks at a fixed area fraction of $\phi=0.5$ and various box lengths. We have used a MATLAB code for simulating systems with $N\leq1600$ disks. For larger sized systems, we have used a custom-written HOOMD-blue~\cite{Anderson2020} script due its faster execution time. In Fig.~\ref{fig:compare_stats}, the mean and variance of the size of the largest cluster computed using HOOMD and MATLAB are compared. The mean cluster size follows the same qualitative trend for both the sets of data, and the difference in numbers could presumably be due to the differing protocols for the implementation of steric-repulsion in the two codes. The peak in the variance, however, is observed at the same frequency for results obtained using the two codes.

\bibliography{supplement}